\documentclass[prd,preprintnumbers,amsmath,amssymb,floatfix]{revtex4}
\usepackage{amsmath,amsfonts,latexsym,amssymb,graphicx,graphics,epsfig,subfigure,color,makeidx,cases}
\usepackage{xcolor,diagbox,enumitem}
\usepackage{multirow}
\usepackage{float}
\usepackage[colorlinks,linkcolor=blue,anchorcolor=blue,citecolor=green,urlcolor=blue]{hyperref}
\usepackage{mathrsfs}

\begin{document}
\title{Real Part Emergence in Purely Imaginary Quasinormal Modes in Perturbed de Sitter Braneworlds}

\author{Hai-Long Jia$^{a}$$^{b}$\footnote{jiahl2024@lzu.edu.cn}}
\author{Wen-Di Guo$^{a}$$^{b}$\footnote{guowd@lzu.edu.cn}}
\author{Yun-Tao Gu$^{a}$$^{b}$\footnote{guyt2024@lzu.edu.cn}}
\author{Yu-Xiao Liu$^{a}$$^{b}$\footnote{liuyx@lzu.edu.cn, corresponding author}}

\affiliation{
$^{a}$Lanzhou Center for Theoretical Physics, 
    Key Laboratory of Theoretical Physics of Gansu Province, 
    Key Laboratory of Quantum Theory and Applications of MoE, 
    Gansu Provincial Research Center for Basic Disciplines of Quantum Physics,
    Lanzhou University, Lanzhou 730000, China \\
$^{b}$Institute of Theoretical Physics \& Research Center of Gravitation,
    School of Physical Science and Technology, 
    Lanzhou University, Lanzhou 730000, China \\
}

\begin{abstract}

    For braneworlds with infinite extra dimensions, an analysis of the stability of the characteristic
    spectrum is essential for understanding their dynamical properties. In this study, we investigate
    the stability of the gravitational quasinormal-mode (QNM) spectrum of a thick de Sitter
    braneworld with a noncompact extra dimension. The unperturbed background
    supports a localized graviton zero mode and a tower of purely imaginary QNMs. We first
    reconstruct the response of the warp factor and tensor effective potential to a smooth
    double-kink deformation of the bulk scalar field, and then introduce a superpotential
    deformation that parametrizes the resulting double-barrier structure. The deformed
    frequency-domain spectrum develops QNM pairs with nonzero real parts. For the
    representative family studied here, a weak deformation can already generate a complex
    branch, although the least-damped pole remains purely imaginary in part of the parameter
    range. In the time domain, the QNM-dominated channel exhibits a transient damped
    oscillatory stage, whereas the original-potential evolution becomes dominated by the
    zero-mode sector at late times over the simulated interval. These results identify the
    double-barrier geometry as the model-specific realization of a more general sensitivity
    of outgoing, non-self-adjoint spectra. They establish a robust model-level mechanism for
    oscillatory QNM imprints and provide a dimensionless benchmark for future source-to-detector
    analyses; quantitative detector detectability and cosmological-parameter inference require
    source, response, and noise modeling beyond the present analysis.

\end{abstract}

\maketitle
\section{Introduction} \label{Introduction}

Since the introduction of the braneworld concept in the 1990s~\cite{Polchinski:1995mt}, 
numerous models~\cite{Arkani-Hamed:1998jmv,Antoniadis:1998ig,Kaloper:1998sw,Randall:1999ee,Randall:1999vf,Kogan:1999wc,Gremm:2000dj,Gregory:2000jc,Dvali:2000rv,Wang:2002pka}
have been proposed as higher-dimensional extensions of general relativity, 
offering new approaches to long-standing issues in gravitational theory. 
Representative examples include the Arkani-Hamed-Dimopoulos-Dvali models~\cite{Arkani-Hamed:1998jmv,Antoniadis:1998ig}
and the Randall-Sundrum-I model~\cite{Randall:1999ee}, 
which address the gauge hierarchy problem through the presence of extra dimensions; 
the Randall-Sundrum-II model~\cite{Randall:1999vf}, 
which first realized the effective localization of gravity in a scenario with an infinite extra dimension; 
and the Dvali-Gabadadze-Porrati model~\cite{Dvali:2000rv}, 
which suggested a possible screening of gravity at the cosmological scale. 
These developments have not only deepened our understanding of higher-dimensional gravity 
but also established the theoretical foundation for exploring observable effects 
that may signal the existence of extra dimensions 
  (for comprehensive reviews, see Refs.~\cite{Dzhunushaliev:2009va,Maartens:2010ar,Liu:2017gcn,Ahluwalia:2022ttu}).

The different extra-dimensional frameworks are not equally suited to the present spectral
question. In ADD-type models, the KK spectrum is primarily set by compactification data,
whereas a localized, continuously deformable scattering potential is not part of the minimal
setup. In thin Randall--Sundrum models, gravity is localized in a distributional geometry and
perturbations must be supplemented by junction conditions. A smooth thick brane generated
by bulk matter instead provides a regular warp factor and tensor potential, a normalizable
graviton zero mode, and a controlled noncompact scattering problem. We therefore use the
thick de Sitter brane as a clean setting in which to separate the general sensitivity of an open
non-self-adjoint QNM problem from the specific cavity mechanism produced by a localized
double-barrier deformation. This choice is intended to isolate the mechanism, not to claim that
the resulting spectral behavior is unique to thick branes.

A distinctive feature of these higher-dimensional theories is that gravitational perturbations contain, 
in addition to the massless zero mode, a tower of massive Kaluza-Klein (KK) excitations. 
The collective spectrum of these modes constitutes the characteristic signature of a given braneworld model, 
with its origins traceable to the early KK framework~\cite{Kaluza:1921tu,klein1926quantum,Overduin:1997sri}. 
In conventional compact models where the extra dimension is topologically an $S^1$,
the KK masses form a discrete spectrum, ensuring consistency with experimental tests of Newtonian gravity. 
The Randall-Sundrum-II scenario, on the other hand, extended the extra dimension to a noncompact $R^1$
and successfully recovered four-dimensional Newtonian gravity through the localization of the gravitational zero mode, 
accompanied by small corrections from the continuum KK sector~\cite{Randall:1999vf}. 
The specific form of these corrections depends sensitively on the effective potential experienced 
by gravitons along the extra dimension. Consequently, different potential profiles lead to distinct 
KK spectral structures and characteristic imprints in the four-dimensional Newtonian potential. 
High-precision measurements of such deviations thus provide a potential avenue for testing 
the existence of extra dimensions 
(see experimental studies in Refs.~\cite{Yang:2012zzb,Tan:2016vwu,Tan:2020vpf,Lee:2020zjt,Ke:2021jtj}).

As braneworld scenarios have evolved, 
investigations of their characteristic spectra have progressed from continuous KK modes to discrete resonant structures. 
In particular, the identification of 
resonant (quasi-bound) modes~\cite{Csaki:2000pp,Brevik:2002yj,Clarkson:2005mg,Melfo:2006hh,Liu:2009ve,Liu:2009uca,Liu:2011wi,Zhong:2016iko,Zhu:2023tzx,Zhu:2024gvl}
and quasinormal modes (QNMs)~\cite{Seahra:2005wk,Seahra:2005iq,Chung:2015mna,Tan:2022vfe,Tan:2023cra,Tan:2024url,Jia:2024pdk,Tan:2024aym,Tan:2024qij,Jia:2024sdk,Deng:2025hfn,E:2025kic,Li:2026dfy}
has enabled a refined characterization of gravitational perturbations, capturing both propagation and dissipation properties. 
This development is closely linked to the advent of gravitational wave astronomy: 
since gravitational waves can propagate into the infinite extra dimension, 
their response on the brane may serve as a potential probe of extra-dimensional effects. 
Consequently, analyzing characteristic modes in gravitational wave signals 
provides a promising means of testing the physical viability of braneworld models, 
in close connection with current observational frontiers~\cite{Konoplya:2023fmh,NANOGrav:2023gor,Koyama:2004cf,Caprini:2018mtu}.
Previous studies~\cite{Seahra:2005iq,Chung:2015mna,Tan:2023cra}
have demonstrated that resonant modes can be interpreted as long-lived QNMs, 
corresponding to massive perturbations with slow decay in the time domain. In general, 
the spectrum of a braneworld model consists of a localized zero mode together with a discrete set of QNMs. 
Depending on the spacetime curvature of the brane, these modes fall into three categories: 
flat branes admit complex QNM frequencies with both real and imaginary parts~\cite{Seahra:2005wk,Tan:2022vfe,Jia:2024pdk}; 
de Sitter (dS) branes yield purely imaginary QNM spectra~\cite{Jia:2024sdk}; 
and anti-de Sitter (AdS) branes exhibit purely real normal modes~\cite{Karch:2000ct,Afonso:2006gi,Liu:2011zy}.

Nevertheless, most existing analyses have been performed in idealized or ``clean'' backgrounds, 
neglecting the inevitable influence of environmental or geometric perturbations. 
This raises a fundamental question: \textit{How stable is the characteristic spectrum of a braneworld model under background perturbations?} 
In our previous work~\cite{Jia:2025saq}, 
we performed a systematic analysis of the spectral stability of QNMs in the flat-brane case. 
The results indicate that the stability is highly sensitive to both the type and the magnitude of perturbations. 
For sufficiently small perturbations or within specific parameter regimes, 
the spectrum remains stable; in contrast, in instability regions, small modulations appear in the time-domain waveform, 
though their amplitudes are typically too weak to be detectable with current detector sensitivities.

Motivated by these findings, the present work concentrates on the spectral stability of the de Sitter thick-brane model. 
Unlike the flat case, the characteristic spectrum of a dS brane consists of purely imaginary QNMs, 
which correspond to non-oscillatory but decaying signals in the time domain.
We aim to investigate whether such purely imaginary modes develop nonvanishing real parts 
when background perturbations are introduced, potentially giving rise to new oscillatory patterns in the waveform. 
Addressing this issue is essential for understanding the dynamical stability of gravitational perturbations 
in higher-dimensional backgrounds and for identifying a model-level waveform
signature that can guide future detector-level assessments.

This question is closely related to, but distinct from, the recent discussion of black-hole QNM spectral
instability. Because QNMs are eigenvalues of an outgoing, non-self-adjoint and generally non-normal
boundary-value problem,
small changes of an effective potential need not produce proportionally small spectral shifts.
Resolvent and pseudospectral studies quantify this sensitivity for black holes~\cite{Jaramillo:2020tuu},
and explicit remote P\"oschl--Teller or Gaussian bumps show how an additional scattering structure can
destabilize even the fundamental mode~\cite{Cheung:2021bol}. Complementary transfer-matrix and
time-domain analyses show that a strongly rearranged spectrum need not imply an equally large
change of the prompt ringdown~\cite{Ianniccari:2024rqs,Jaramillo:2021gw,Berti:2022xfj}. Our analysis
is a direct smooth-deformation test rather than a calculation of the full pseudospectrum. It isolates
how this general open-system sensitivity is realized in a dS braneworld, where the unperturbed QNMs
lie on the imaginary axis and a matter-induced double-barrier geometry supplies the specific cavity
mechanism.

The paper is organized as follows.
In Sec.~\ref{dS braneworld}, we review the dS braneworld model and 
discuss in detail how background perturbations affect the brane energy density distribution and the associated effective potential.
In Sec.~\ref{Stability analysis of QNMs}, we present an explicit parametrization of the perturbations arising in the dS braneworld. 
We then investigate their impact on the QNM spectrum and on the time-domain evolution, 
followed by a detailed comparison and physical interpretation of the results.
Finally, a brief summary and discussion are provided in Sec.~\ref{conclusion}. 
In the following, capital Latin indices $M,N,\cdots = 0,1,2,3,5$ are used for the five-dimensional bulk coordinates, 
Greek indices $\mu,\nu,\cdots = 0,1,2,3$ for the four-dimensional spacetime, 
and Latin indices $i,j,\cdots = 1,2,3$ for the three-dimensional space.

\section{The de Sitter Braneworld} \label{dS braneworld}

\subsection{Review of a thick de Sitter braneworld} \label{thick dS brane}

We begin by considering a five-dimensional thick dS brane generated by a single scalar field $\phi(z)$. 
The line element takes the form~\cite{Randall:1999ee,Randall:1999vf}
\begin{equation}%
    \label{line-element}
    ds^2 = e^{2A(z)}\left(\gamma_{\mu\nu}dx^{\mu}dx^{\nu}+dz^2\right), 
\end{equation}
where $z$ represents the conformal coordinate of the extra dimension 
and $x^{\mu}$ are the standard four-dimensional spacetime coordinates. 
In this expression, $A(z)$ is the warp factor, and 
$\gamma_{\mu\nu}=\mathrm{diag}\left\{-1,e^{2\alpha t},e^{2\alpha t},e^{2\alpha t}\right\}$
denotes the four-dimensional dS metric, with $\alpha >0$ associated with the Hubble constant of the dS spacetime.
The action is~\cite{DeWolfe:1999cp,Gremm:1999pj,Csaki:2000fc}
\begin{equation}%
    \label{action}
    S = \int d^4x dz \sqrt{-g} \left(\frac{1}{2} R - \frac{1}{2}\nabla_{M}\phi \nabla^{M}\phi - V(\phi) \right).
\end{equation}
Variation of the action~\eqref{action} yields the five-dimensional Einstein equations 
together with the scalar field equation. 
Substituting the explicit form of the metric~\eqref{line-element} into these equations 
and analyzing them, we find that only two of the resulting equations are independent,
\begin{align}%
    \label{independent-equation-1}
    3A'' + 3A'^2 -3\alpha^2  &= -\frac{1}{2}\phi'^2-e^{2A} V(\phi),\\
    \label{independent-equation-2}
    6A'^2-6\alpha^2 &= \frac{1}{2}\phi'^2-e^{2A} V(\phi),
\end{align}
where the prime denotes the derivative with respect to $z$,
and there are three unknown functions to be determined. 
Accordingly, an additional condition is generally required to obtain the complete set of solutions.
In particular, by combining Eqs.~\eqref{independent-equation-1} and~\eqref{independent-equation-2} 
we derive the relation between $A(z)$ and $\phi(z)$:
\begin{equation}%
    \label{relation-A-phi}
    3A'' - 3A'^2 + 3\alpha^2  = -\phi'^2. 
\end{equation}

Next, we turn to the discussion of tensor perturbations of the metric in the thick dS brane.
Specifically, we consider a perturbation $h_{ij}(x^{\lambda},z)$ satisfying the transverse-traceless condition 
$\delta^{ij}h_{ij}=0=\delta^{ij}\partial_{i}h_{jk}$, 
for which the perturbed metric can be written as 
\begin{equation}%
    \label{line-element-perturbation}
    ds^2=e^{2A(z)}\left[-dt^2+ e^{2\alpha t}\left(\delta_{ij}+h_{ij}\right)dx^{i}dx^{j}+dz^2\right].
\end{equation}
This type of perturbation has been discussed in detail in our previous work~\cite{Jia:2024sdk},
and essentially describes perturbations on the three-dimensional spatial section of the brane.
By substituting Eq.~\eqref{line-element-perturbation} into the Einstein equations,
we derive the perturbation equation for $h_{ij}$, namely
\begin{equation}%
    \label{five-wave-equation-hij}
    \left[-\left(\partial_{t}^{2}+3\alpha \partial_{t}\right) + e^{-2\alpha t} \partial_{k}\partial^{k} + \partial_{z}^{2} + 3 \left(\partial_{z} A(z)\right)\partial_{z} \right] h_{ij} = 0.
\end{equation}
Guided by separation of variables and by the structure of the perturbation equation, we assume that the perturbation $h_{ij}$ takes the form
\begin{equation}%
    \label{hij-assumption}
    h_{ij}=\varepsilon_{ij} e^{-ip_{k}x^{k}} \Phi(t,z), \quad \Phi(t,z) = e^{-\frac{3}{2}\alpha t} e^{-\frac{3}{2}A(z)} \Psi(t,z).
\end{equation}
The perturbation equation~\eqref{five-wave-equation-hij} can be further reduced to 
\begin{equation}%
    \label{psi-equation}
    \left[-\partial_{t}^{2} + \frac{9}{4} \alpha ^2 - e^{-2 \alpha t} p^2  + \partial_{z}^{2}  -V(z) \right] \Psi (t,z) =0, 
\end{equation}
where $p^2=\delta^{ij}p_{i}p_{j}$, and 
\begin{equation}%
    \label{effective-potential}
    V(z) = \frac{3}{2}A'' + \frac{9}{4} A'^2
\end{equation}
denotes the effective potential for the tensor perturbation.
By comparing the tensor perturbation equation in the dS brane with that in the flat brane scenario,
we find that the presence of the $e^{-2 \alpha t} p^2$ term 
precludes the application of a Fourier transform to the time component,
and therefore the equation cannot be directly recast into the standard Schr\"odinger-like form.
However, it is observed that this term decays exponentially with time and has a negligible effect at late times.
Accordingly, we divide the analysis into two distinct cases: 
\begin{itemize}
    \item $p^2 \neq 0$: We employ numerical methods to solve the equation~\eqref{psi-equation}.
    In our previous work~\cite{Jia:2024sdk}, we found that this term affects primarily the early-time behavior of the signal,
    and in the limit $e^{-2 \alpha (t-t_0)} p^2 \ll 1$, the system effectively reduces to the $p^2 = 0$ case. 
    \item $p^2 = 0$: In this case, the perturbation equation~\eqref{psi-equation} further simplifies to 
    \begin{equation}%
        \label{psi-equation-p0}
        \left[-\partial_{t}^{2}  + \partial_{z}^{2}  -V_{\text{re}}(z) \right] \Psi (t,z) =0,   
\end{equation}
    where $V_{\text{re}}(z)= V(z) - 9\alpha ^2 / 4$ denotes the reduced effective potential.
    Then we can solve the equation in the frequency domain by assuming $\Psi (t,z) \sim e^{-i\omega t} \psi(z)$,
    which leads to the Schr\"odinger-like equation 
    \begin{equation}%
        \label{extra-dimensional-equation-re}
        \left( -\partial_z^2 + V_{\text{re}}(z) \right) \psi(z) = \omega^2 \psi(z)  
\end{equation}
    in the dS braneworld. A detailed discussion of this procedure can be found in Ref.~\cite{Jia:2024sdk}.
\end{itemize}

By specifying the explicit form of the interaction potential $V(\phi)$ for the background scalar field,
Wang (2002) proposed a class of thick dS brane models~\cite{Wang:2002pka}.
The complete background solution reads
\begin{align}%
    \label{solution-Az}
    A_0(z) &=-n\ln \left[\cosh\left(\beta z\right)\right], \\
    \label{solution-phi}
    \phi_0(z) &=\sqrt{3n(1-n)} \sin^{-1}\left[\tanh\left(\beta z\right)\right],\\ 
    \label{solution-Vphi}
    V(\phi_0) &=n\beta^2 \frac{3(1+3n)}{2} \cos^{2(1-n)}\left(\frac{\phi_0}{\sqrt{3n(1-n)}}\right), 
\end{align}
where $0<n<1$ and $\beta >0$ are arbitrary constants,
and their relation to $\alpha$ is expressed as $\alpha^2=n^2 \beta^2$. 
In our previous work~\cite{Jia:2024sdk},
we carried out a detailed study of the QNM spectrum of the tensor perturbation in this model,
which comprises a zero mode together with a discrete set of purely imaginary modes.

In the present paper, we adopt this thick dS brane model as a representative case
to investigate the stability of the purely imaginary QNM spectrum in the dS braneworld.

\subsection{The perturbation mechanism} \label{mechanism}

In our previous work~\cite{Jia:2025saq}, 
we verified that perturbations of the background field do induce deformations in the effective potential,
which in turn influence the characteristic spectrum in the flat brane background.
In the following, we extend this analysis to the dS brane case,
where a double-kink perturbation is introduced into the original background scalar field,
leading to a symmetric deformation of the warp factor. 
Furthermore, we discuss how such matter field perturbations influence the effective potential.

The double-kink profile is chosen as a physically motivated, smooth, and controllable
matter-sector perturbation, not as a claim of uniqueness. Kink scalar configurations are standard
smooth realizations of domain walls and thick branes~\cite{DeWolfe:1999cp,Gremm:1999pj}, and
double-kink profiles occur explicitly in deformed or split-brane
solutions~\cite{Liu:2011ysa,Guo:2011qtj}. It is therefore reasonable to use such a configuration
as a perturbation of the background matter field. The deformation in Eq.~\eqref{phi-kink-perturbation}
has odd parity, namely, it is an odd function of the extra-dimensional coordinate $z$; consequently, the reconstructed warp factor and tensor effective potential remain even,
while its strength and separation can be varied independently through the amplitude and position
parameters. Our strategy is first to reconstruct numerically how this background perturbation
deforms the tensor potential and then to parameterize the resulting deformation class for the
subsequent spectral analysis. Other smooth matter profiles can be treated by the same procedure,
and they are expected to exhibit the same cavity-type QNM mechanism whenever their backreaction
produces two sufficiently separated and reflective structures in the tensor potential.

\begin{figure*}[htb]
    \begin{center}
    \subfigure[~Bulk scalar field ($\epsilon_m=0.1$)]  {\label{perturbated-phiz}
    \includegraphics[width=5.6cm]{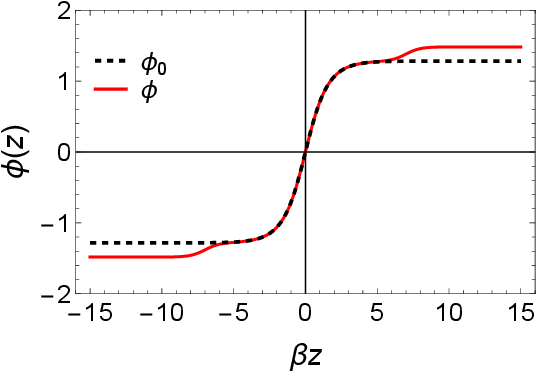}}
    \subfigure[~Warp factor ($\epsilon_m=1$)]  {\label{perturbated-Az}
    \includegraphics[width=5.6cm]{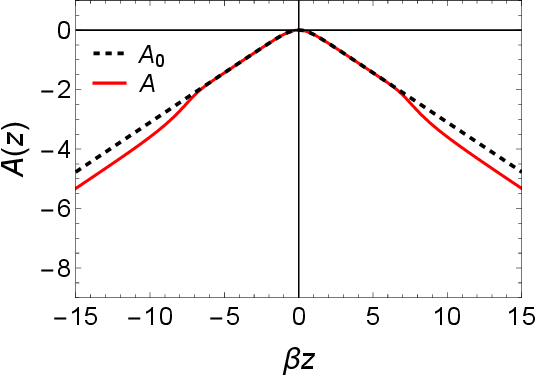}}\\
    \subfigure[~Effective potential ($\epsilon_m=0.5$)]  {\label{perturbated-Vz}
    \includegraphics[width=5.6cm]{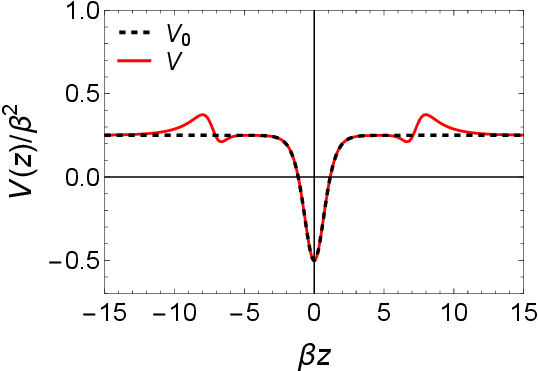}}
    \subfigure[~Energy density ($\epsilon_m=0.1$)]  {\label{perturbated-EDz}
    \includegraphics[width=5.6cm]{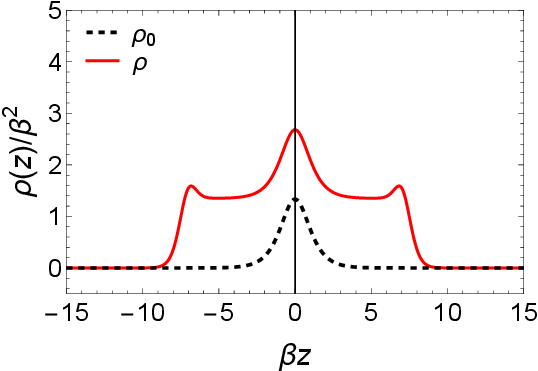}}
    \end{center}
    \caption{Deformations of the bulk scalar field, the warp factor, the effective potential,
    and the energy density induced by the matter double-kink perturbation. 
    Here $n=1/3$ and $a_m=7/\beta$. 
    Note that the comparatively large perturbation amplitude is adopted 
    solely to clearly illustrate the influence of matter perturbations 
    on the relevant physical quantities. }
    \label{figure-1}
\end{figure*}

We assume that the background scalar field 
supports a double-kink perturbation (see Fig.~\ref{perturbated-phiz}), 
\begin{equation}%
    \label{phi-kink-perturbation}
    \phi(z) = \phi_0(z) + \epsilon_m \left\{\tanh\left[\beta (z-a_m)\right] + \tanh\left[\beta (z+a_m)\right]\right\},
\end{equation}

where the parameters $\epsilon_m$ and $a_m$ parameterize 
the amplitude and location of the matter perturbation, respectively.
The kink width is fixed by the inverse scale $\beta^{-1}$ in this scan.

Such a deformation in the background matter field alters the background solution,
while leaving the form of the field equations unchanged. 
Consequently, the perturbed warp factor (see Fig.~\ref{perturbated-Az}) 
can still be derived from Eq.~\eqref{relation-A-phi}.
Since Eq.~\eqref{relation-A-phi} cannot be solved analytically in general, 
we perform a numerical integration subject to the conditions $A(0)=0$ and $A'(0)=0$.
Accordingly, using Eq.~\eqref{effective-potential} and 
the expression for the energy density measured by a static observer $u^{M}=(e^{-A(z)},0,0,0,0)$ in the bulk,
\begin{equation}%
    \label{energy-density}
    \rho(z) = T_{MN}u^{M}u^{N} = - e^{-2A(z)} \left(3A'' + 3A'^2- 3\alpha^2 \right),
\end{equation} 
we derive the perturbed effective potential (see Fig.~\ref{perturbated-Vz}) 
and the corresponding modification of the energy density profile (see Fig.~\ref{perturbated-EDz}).

In Fig.~\ref{figure-1}, the black dashed and red solid curves represent the characteristic behaviors 
of the physical quantities without and with the perturbation, respectively, with $n=1/3$. 
As illustrated in Fig.~\ref{perturbated-Az}, under the perturbation of the background scalar field, 
the warp factor displays a downward deviation at the perturbation
positions $z=\pm a_m=\pm 7/\beta$,
and rapidly recovers, approaching the same asymptotic slope as the unperturbed configuration.
This behavior differs from that observed in a flat brane background, 
where the perturbation induces a slope deviation of the warp factor in distant regions~\cite{Jia:2025saq}.
In contrast, the change in the brane energy density is more pronounced, 
exhibiting a plateau-like uplift within the
dimensionless range $\beta z\in(-\beta a_m,\beta a_m)$.
Moreover, our numerical results indicate that perturbations 
of the background matter field significantly modify the asymptotic structure 
of the bulk geometry along the extra dimension. 
While the unperturbed dS brane asymptotically approaches a five-dimensional Minkowski spacetime at large $|z|$, 
the perturbed configuration tends toward an asymptotically AdS spacetime, 
because the perturbation shifts the asymptotic value of the bulk scalar potential to a negative constant, 
thereby generating a negative effective cosmological constant.
Such a transition arises whenever the perturbation is introduced, 
and the negative cosmological constant of the resulting AdS geometry 
is entirely determined by the perturbation parameters $\epsilon_m$ and $a_m$.
Regarding the effective potential, two bumps emerge at the perturbation positions $(z=\pm a_m)$,
forming a double-barrier-like structure.
This structure has a significant influence on the spectral properties and may give rise to the appearance of resonant modes.

The physical ingredient relevant to the spectral pattern found below is therefore the resulting
cavity geometry, rather than the hyperbolic-tangent form of the scalar perturbation itself. Two
separated reflective structures form a leaky scattering cavity. Standard double-barrier and echo
analyses show that its resonances are fixed by round-trip phase accumulation and leakage, with the
leading real-frequency spacing given by
$\Delta\operatorname{Re}(\omega)\simeq\pi/L$
\cite{Ianniccari:2024rqs,Cardoso:2016rao,Mark:2017dnq,Tan:2024qij,Jia:2025saq}. For structures near
$z=\pm a$, the effective length is $L\simeq2a$, so this reduces to the spacing estimate tested in
Sec.~\ref{Frequency-domain}. Thus another localized profile with comparable barrier separation and
reflectivity should also generate complex cavity resonances, although its damping rates and detailed
spectral trajectories will depend on the profile.

This statement has a definite scope. A localized perturbation that does not form an effective
cavity need not display the same approximately equally spaced spectrum. It can nevertheless move
QNM poles away from the imaginary axis and produce complex frequencies with nonzero real parts;
that behavior is then governed by the specific potential deformation rather than by the present
cavity quantization mechanism~\cite{Cheung:2021bol,Jia:2025saq}. The method itself remains
applicable: one can reconstruct the effective-potential response to another background perturbation,
parameterize that response, and repeat the frequency- and time-domain analyses.

\begin{figure*}[htb]
    \begin{center}
    \subfigure[~Energy density ($n=1/3,a_m=7/\beta$)]  {\label{perturbated-deltaEDz-2}
    \includegraphics[width=5.6cm]{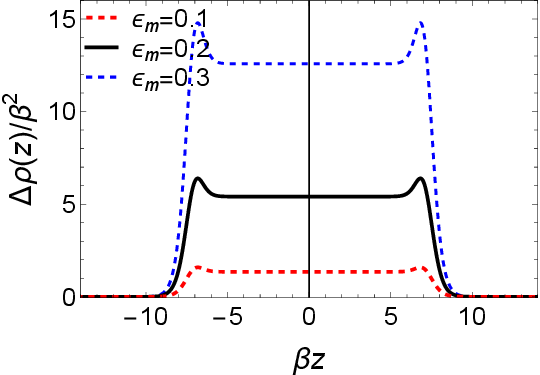}}
    \subfigure[~Energy density ($n=1/3,\epsilon_m=0.1$)]  {\label{perturbated-deltaEDz-1}
    \includegraphics[width=5.6cm]{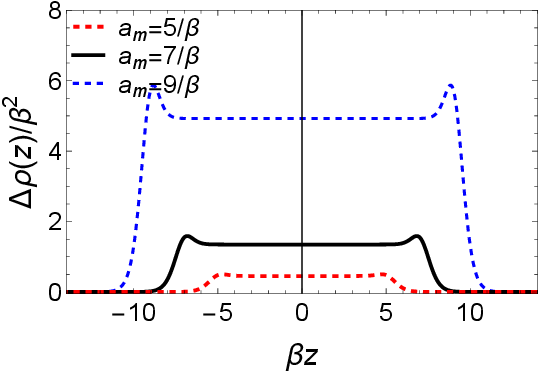}}
    \subfigure[~Energy density ($\epsilon_m=0.1,a_m=7/\beta$)]  {\label{perturbated-deltaEDz-3}
    \includegraphics[width=5.6cm]{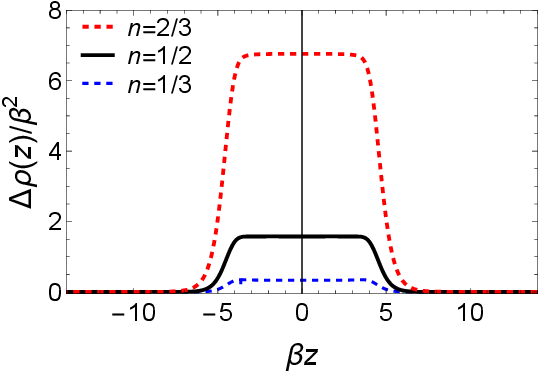}}\\
    \subfigure[~Effective potential ($n=1/3,a_m=7/\beta$)]  {\label{perturbated-deltaVz-2}
    \includegraphics[width=5.6cm]{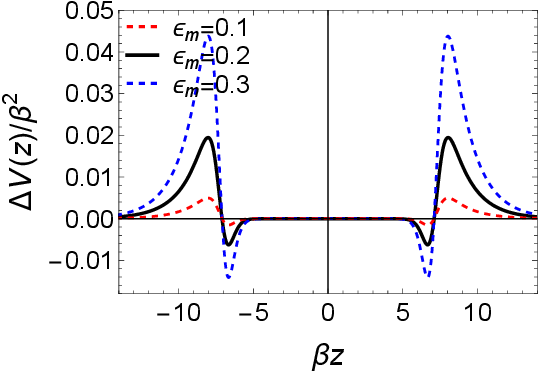}}
    \subfigure[~Effective potential ($n=1/3,\epsilon_m=0.1$)]  {\label{perturbated-deltaVz-1}
    \includegraphics[width=5.6cm]{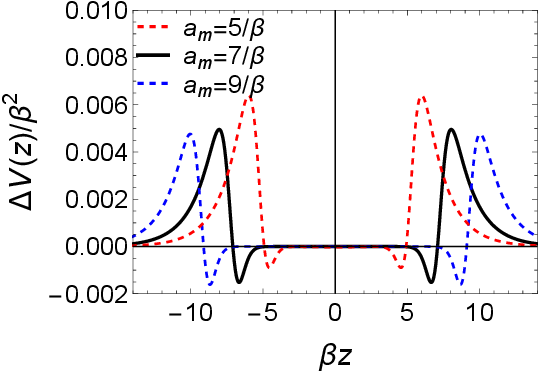}}
    \subfigure[~Effective potential ($\epsilon_m=0.1,a_m=7/\beta$)]  {\label{perturbated-deltaVz-3}
    \includegraphics[width=5.6cm]{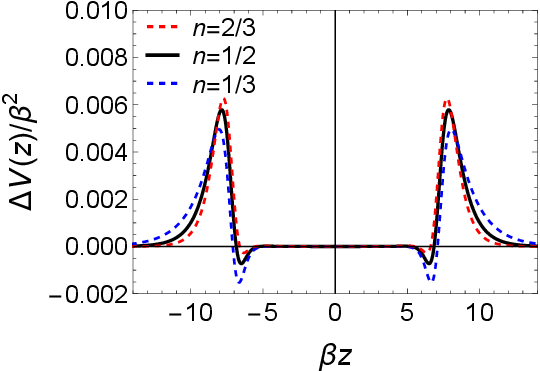}}
    \end{center}
    \caption{Deformations of the energy density and the effective potential induced by the matter kink-type perturbation 
    with respect to the parameters $\epsilon_m$, $a_m$, and $n$.}
    \label{figure-2}
\end{figure*}

\begin{figure*}[htb]
    \begin{center}
    \subfigure[~Energy density ($a_m=7/\beta$)]  {\label{perturbated-deltarhoz-em}
    \includegraphics[width=5.6cm]{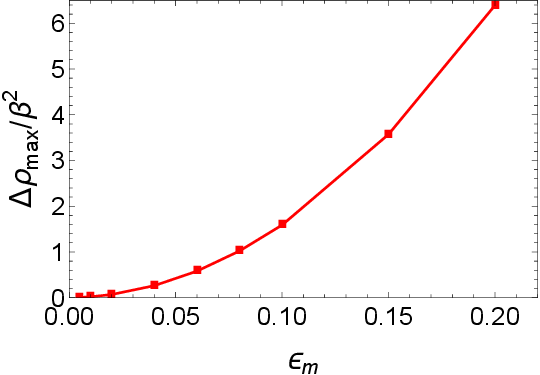}}
    \subfigure[~Energy density ($\epsilon_m=0.1$)]  {\label{perturbated-deltarhoz-am}
    \includegraphics[width=5.6cm]{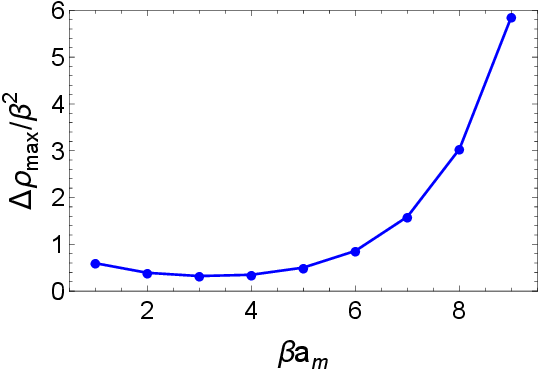}}\\
    \subfigure[~Effective potential ($a_m=7/\beta$)]  {\label{perturbated-deltaVz-em}
    \includegraphics[width=5.6cm]{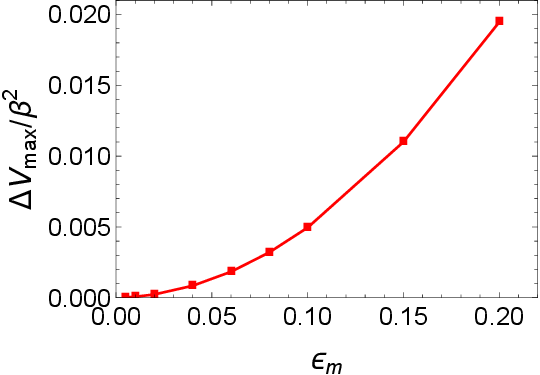}}
    \subfigure[~Effective potential ($\epsilon_m=0.1$)]  {\label{perturbated-deltaVz-am}
    \includegraphics[width=5.6cm]{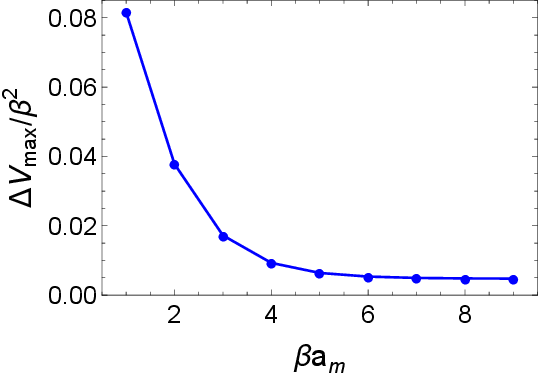}}
    \end{center}
    \caption{Peak values of the variations of the energy density and the effective potential 
    as functions of the perturbation parameters $\epsilon_m$ and $a_m$, with $n=1/3$.}
    \label{figure-3}
\end{figure*}

To obtain a more precise understanding of how variations in the brane energy density 
influence the effective potential, we perform a systematic analysis 
by classifying the perturbation parameters $\epsilon_m$ and $a_m$,
as well as the solution family parameter $n$ in Fig.~\ref{figure-2}. 
We define the variations as $\Delta\rho=\rho-\rho_0$ and $\Delta V=V-V_0$.
Specifically, we compute and display the peak values of these variations 
as functions of $\epsilon_m$ and $a_m$ in Fig.~\ref{figure-3}. 
The results can be summarized as follows:
\begin{itemize}
    \item Effect of the perturbation amplitude $\epsilon_m$: \\
    As shown in Figs.~\ref{perturbated-deltaEDz-2} and~\ref{perturbated-deltaVz-2}, 
    as $\epsilon_m$ increases proportionally, the variations 
    of the energy density and effective potential do not exhibit linear scaling.
    From Figs.~\ref{perturbated-deltarhoz-em} and~\ref{perturbated-deltaVz-em}, 
    the scalings of both quantities exhibit similar behaviors.
    Our numerical analysis indicates the following relation between their peak magnitudes:
    \begin{equation}%
        \label{Vrho-relation}
        \frac{\max_z|\Delta V(z)|}{\max_z|\Delta\rho(z)|}
        \simeq 3\times10^{-3}=0.3\%, \qquad a_m=7/\beta,
\end{equation}
    where the coefficient of proportionality depends on the perturbation position: 
    the larger the perturbation distance, the smaller the ratio.
    \item Effect of the perturbation position $a_m$: \\
    As shown in Figs.~\ref{perturbated-deltaEDz-1} and~\ref{perturbated-deltarhoz-am}, 
    the variation of the energy density increases with $a_m$,
    in contrast to the flat brane case.
    This behavior can be attributed to the modification of the asymptotic structure 
    of the brane along the extra dimension, 
    which enhances the energy density distribution at large distances.
    By contrast, the variation of the effective potential decreases with $a_m$ 
    (see Figs.~\ref{perturbated-deltaVz-1} and~\ref{perturbated-deltaVz-am}).
    Moreover, although the parameters are varied proportionally, 
    the corresponding quantities exhibit an approximately exponential dependence on the parameters.
    \item Dependence on the solution family parameter $n$: \\
    The variation of the energy density differs markedly for different solution families, 
    whereas the corresponding deviations in the effective potential are comparatively small.
    As $n$ increases, the potential barriers at the perturbation positions become increasingly pronounced, 
    while the potential wells become shallower.
    \item General characteristics: \\
    Regardless of the parameter choices, the variation in the energy density consistently displays a plateau-like uplift, 
    often accompanied by a small peak near the perturbation position within a certain parameter range.
    In contrast, the deformation of the effective potential consistently resembles a double-barrier structure, 
    with the barrier separation approximately equal to twice the perturbation position.
\end{itemize}

\begin{figure*}[htb]
    \begin{center}
    \subfigure[~($a_m=7/\beta$)]  {\label{deltaVzvsdeltarhoz-em}
    \includegraphics[width=5.6cm]{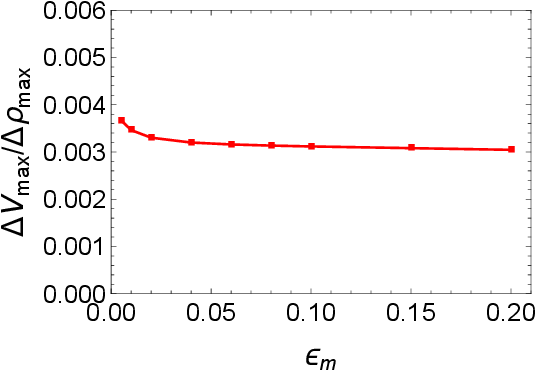}}
    \subfigure[~($\epsilon_m=0.1$)]  {\label{deltaVzvsdeltarhoz-am}
    \includegraphics[width=5.6cm]{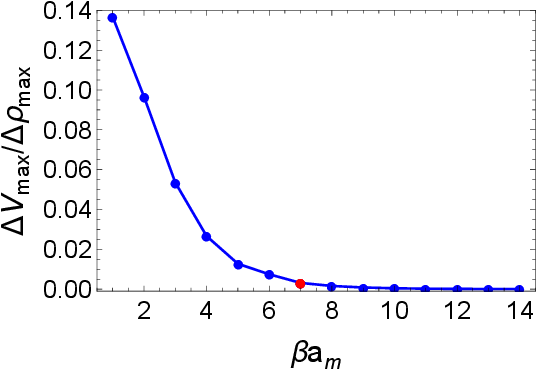}}
    \end{center}
    \caption{Ratio of the peak variation in the effective potential to that in the energy density 
    as a function of the perturbation parameters $\epsilon_m$ and $a_m$, with $n=1/3$. 
    The red dot in the right panel is at $a_m=7/\beta$ and corresponds to the point at $\epsilon_m=0.1$ in the left panel.}
    \label{figure-4}
\end{figure*}

From Eq.~\eqref{Vrho-relation}, the matter-to-tensor response is strongly
attenuated in this representative scan: even an energy-density rearrangement comparable to
the background scale can correspond to a much smaller, but not identically vanishing, tensor
effective-potential deformation. As illustrated in Fig.~\ref{figure-4}, 
the influence of the perturbation amplitude on this relation is significantly weaker than that of the perturbation position.
Figure~\ref{deltaVzvsdeltarhoz-am} further illustrates that perturbations occurring near the brane 
exert a stronger impact than those occurring at larger distances. When the parameter $a_m$ approaches zero, 
perturbations that produce an energy density variation one order of magnitude smaller than 
the background still produce a noticeable effect on the effective potential.
Accordingly, this relation will be used in the subsequent evaluation of the stability of the QNM spectra.

In the flat-brane setting, Ref.~\cite{Jia:2025saq} showed that suitable
kink- and soliton-type background-field deformations can preserve gravitational zero-mode
localization. Motivated by that result, below we verify the corresponding statement explicitly
for the present dS superpotential deformation: its bounded primitive leaves the asymptotic
zero-mode falloff, and hence normalizability, unchanged.
By contrast, the QNMs are expected to be sensitive to such perturbations, 
and the corresponding modifications in their spectra will be discussed in detail in the following section.

\section{Stability analysis of QNMs} \label{Stability analysis of QNMs}

In the previous section, we reviewed several fundamental features of the dS braneworld 
and introduced perturbations by incorporating a kink-type function into the background scalar field 
to model possible disturbances in the system.
By analyzing the ratio of the peak deformation of the effective potential to that of the energy density, 
we characterized the sensitivity of the potential to such perturbations.
Our analysis indicated that the position of the perturbation 
has a stronger impact on this ratio than its amplitude: 
perturbations occurring near the brane induce more pronounced modifications in the effective potential 
than those farther away (see Fig.~\ref{deltaVzvsdeltarhoz-am}).
It is therefore important to determine how these perturbation-induced
deformations affect the QNM spectrum and the dimensionless time-domain response before
drawing any detector-level conclusion for gravitational signals in the dS braneworld.

Before investigating how deformations of the effective potential influence the QNM spectrum,
we first introduce an explicit and tractable parametrization for such deformations.
Since an exact analytic expression for the deformation is generally unavailable and realistic
perturbations may take various forms, it is appropriate to adopt a parametrized description that
captures the essential behavior and permits a systematic analysis.

Our procedure separates the physical reconstruction from the efficient spectral scan. The
Einstein--scalar calculation in Sec.~\ref{mechanism} first identifies how a background matter
perturbation deforms the warp factor and tensor potential. We then parameterize that response at
the superpotential level rather than add an arbitrary function directly to $V$. This is a
structural choice: in an Einstein--scalar braneworld, $W=-3A'/2$, while
$V_-=W^2-W'$ and $V_+=W^2+W'$ denote the original and dual tensor potentials, respectively;
the tensor potential is not an independent background input
\cite{DeWolfe:1999cp,Gremm:1999pj,Csaki:2000fc,Jia:2025saq}. A generic additive $\delta V$
need not admit a globally regular superpotential, preserve the factorized tensor operator, or
retain a normalizable zero-energy state. By specifying $W=W_0+\delta W$, the full original
operator remains
\begin{equation}%
 \mathcal{H}_-^{\mathrm{full}}
 =-\partial_z^2+V_-^{\mathrm{full}}
 =(-\partial_z+W)(\partial_z+W),\qquad
 \psi_{0,-}^{\mathrm{full}}(z)
 =\mathcal{N}\exp\!\left[-\int_0^z W(s)\,ds\right].
 \label{full-zero-mode}
\end{equation}
For the present background, $W_0=(3n\beta/2)\tanh(\beta z)$. The parametrization introduced
below in Eq.~\eqref{parametrized-form} has the bounded primitive
\begin{equation}%
 \Delta S(z)\equiv\int_0^z\delta W(s)\,ds
 =\frac{\epsilon}{2}\left\{
 \tanh[\beta(z-a)]-\tanh[\beta(z+a)]+2\tanh(\beta a)\right\}.
 \label{bounded-primitive}
\end{equation}
Consequently,
$\psi_{0,-}^{\mathrm{full}}=\mathcal{N}\operatorname{sech}^{3n/2}(\beta z)e^{-\Delta S(z)}
\sim e^{-3n\beta|z|/2}$ and is square integrable for $n>0$. Thus the full deformation retains
the localized graviton zero mode. This localization-preserving use of a superpotential is the
same structural strategy employed for the flat-brane case in Ref.~\cite{Jia:2025saq}.

The analytic $\delta W$ should nevertheless be read as an effective-potential proxy motivated by
the numerical Einstein--scalar response; specifying it does not prove a unique closed-form
Einstein--scalar completion. Apart from the full-potential cross-check discussed below, the
reported parameter scans and time-domain evolutions use
$V_\pm^{(1)}=V_\pm^{\mathrm{full}}-(\delta W)^2$, with the notation above. The corresponding truncated operators are not exactly
factorized: the truncated intertwining relation and zero-mode equation agree with the
corresponding full equations through $\mathcal{O}(\epsilon)$, with residuals of
$\mathcal{O}(\epsilon^2)$. This perturbative order counting is not a uniform QNM-error bound;
all exact factorization, isospectrality, and zero-mode statements below refer to the full
original--dual pair.

Following the standard supersymmetric-quantum-mechanics factorization
convention~\cite{Cooper:1994eh},
the effective potential~\eqref{effective-potential} can be factorized 
in terms of a superpotential $W(z)$ as
\begin{equation}%
    V(z)= W^2(z)- W'(z).
\end{equation}
We next introduce a small perturbation $\delta W(z;\epsilon,a)$
added to the background superpotential $W_0(z)$,
thereby generating the corresponding perturbation in the effective potential, i.e.,
\begin{equation}%
    W(z)=W_0(z)+ \delta W(z), 
\end{equation}

For the numerical Einstein--scalar background constructed in Sec.~\ref{mechanism}, the
superpotential is fixed directly by the reconstructed warp factor:
\begin{equation}%
 W_{\mathrm{ES}}(z)=-\frac{3}{2}A'(z),\qquad
 \delta W_{\mathrm{ES}}(z)=-\frac{3}{2}\big[A'(z)-A_0'(z)\big].
 \label{ES-superpotential}
\end{equation}
Its exact tensor-potential response is therefore
\begin{equation}%
 \delta V_{\mathrm{ES}}
 =2W_0\delta W_{\mathrm{ES}}-\delta W_{\mathrm{ES}}'
 +(\delta W_{\mathrm{ES}})^2.
 \label{ES-potential-response}
\end{equation}
Equations~\eqref{ES-superpotential} and~\eqref{ES-potential-response} are the precise bridge
between the self-consistent background and its tensor scattering potential.

The parametrized superpotential deformation results in the perturbed potential
\begin{equation}%
    \label{perturbed-effective-potential}
    V(z) = V_0(z) +  \delta V(z;\epsilon,a),
\end{equation}
where
\begin{align}%
    V_0(z) &= W_0^2(z)- W_0'(z)= \frac{9\beta^2 n^2}{4}-\frac{(9n^2+6n)\beta^2}{4}\operatorname{sech}^2(\beta z),\\
    \delta V_{\mathrm{full}}(z)
    &\;=2W_0(z)\delta W(z)-\delta W'(z)+[\delta W(z)]^2,\notag\\
    \delta V^{(1)}(z)
    &\;=2W_0(z)\delta W(z)-\delta W'(z).
\end{align}

The same quadratic contribution $[\delta W(z)]^2$ appears in both the full original and dual
potentials and is generated exactly by expanding $W^2=(W_0+\delta W)^2$. Since
$\delta W=\mathcal{O}(\epsilon)$, this is the quadratic $\mathcal{O}(\epsilon^2)$ contribution;
the first-order model $\delta V^{(1)}$ is defined by omitting only this term.

It is useful to distinguish the exact original and dual potentials and their first-order
expansions:
\begin{equation}%
\begin{aligned}
 V_-^{\mathrm{full}}&=W^2-W',&
 V_+^{\mathrm{full}}&=W^2+W',\\
 V_-^{(1)}&=W_0^2-W_0'+2W_0\delta W-\delta W',&
 V_+^{(1)}&=W_0^2+W_0'+2W_0\delta W+\delta W'.
\end{aligned}
\label{first-order-partners}
\end{equation}
The frequency-domain data in Figs.~\ref{figure-6} and~\ref{figure-7} were generated with
$V_+^{(1)}$; its nonzero modes are used as perturbative proxies for those of the original
$V_-^{(1)}$ channel. The time-domain calculations use $V_-^{(1)}$ and $V_+^{(1)}$ for the
original and dual evolutions, respectively. The full factorized original and dual operators
obey an exact intertwining relation, whereas removing their common $[\delta W]^2$ term leaves
an $\mathcal{O}(\epsilon^2)$ residual in the corresponding truncated relation. This operator
statement is not a uniform bound on individual QNM shifts; exact isospectrality is lost, and
the representative mismatch is quantified below.

The omitted contribution is quantitatively controlled for the parameter range studied here.
Equation~\eqref{parametrized-form} implies
$\lVert\delta W\rVert_\infty\leq\epsilon\beta/2$, where
$\lVert f\rVert_\infty\equiv\sup_{z\in\mathbb{R}}|f(z)|$ is the supremum norm, and hence
\begin{equation}%
 \lVert(\delta W)^2\rVert_\infty\leq \frac{\epsilon^2\beta^2}{4}.
 \label{quadratic-bound}
\end{equation}
For the largest amplitude used in the numerical scans, $\epsilon=0.1$, and for
$n=1/3$ and $2\leq\beta a\leq8$, direct maximization gives
\begin{equation}%
 \frac{\lVert(\delta W)^2\rVert_\infty}
 {\lVert2W_0\delta W-\delta W'\rVert_\infty}<3.4\times10^{-2},
 \qquad
 \frac{\lVert(\delta W)^2\rVert_\infty}{\lVert V_0\rVert_\infty}
 \leq5.0\times10^{-3}.
 \label{quadratic-ratios}
\end{equation}
The first ratio measures the largest omitted quadratic contribution relative to the retained
first-order deformation, while the second compares it with the scale of the unperturbed
potential. Thus, even at the largest scanned amplitude, the omitted term is below $3.4\%$
of the retained deformation and below $0.5\%$ of the background-potential scale.
As a representative check at the largest amplitude, we directly integrated the wave equation for all four potentials
at $(n,\epsilon,\beta a)=(1/3,0.1,5)$ using the outgoing-wavefunction shooting procedure
employed in the frequency-domain calculation. The QNM frequency computed with the first-order
dual potential $V_+^{(1)}$ is
$\omega_{1,+}^{(1)}/\beta=0.326814-0.152781i$. The corresponding frequency computed with
the first-order original potential $V_-^{(1)}$ is
$\omega_{1,-}^{(1)}/\beta=0.324204-0.151172i$. The relative difference between these two
truncated-potential frequencies is $8.50\times10^{-3}$ and explicitly measures the loss of
exact isospectrality at this amplitude. Restoring the full potentials gives the same
full-potential frequency in the original and dual calculations,
\begin{equation}%
 \omega_{1,-}^{\mathrm{full}}/\beta
 =\omega_{1,+}^{\mathrm{full}}/\beta
 =0.326232-0.149741i,
 \label{full-potential-mode}
\end{equation}
with the original- and dual-potential calculations agreeing within $2.2\times10^{-12}$.
Relative to the corresponding $V_+^{(1)}$ and $V_-^{(1)}$ roots, the relative
complex-frequency shifts
$|\omega_{1,\pm}^{\mathrm{full}}-\omega_{1,\pm}^{(1)}|/
|\omega_{1,\pm}^{(1)}|$ are $8.58\times10^{-3}$ and $6.94\times10^{-3}$,
respectively. All four calculations retain a clearly nonzero real part.
For $\epsilon\leq0.01$, the first ratio in Eq.~\eqref{quadratic-ratios} is below
$3.4\times10^{-3}$. At the largest scanned amplitude, restoring the quadratic term changes
this representative complex frequency by less than one percent, while its nonzero real part
persists. The omitted term therefore does not alter the central qualitative conclusion.
Quantitatively complete QNM values at $\epsilon=0.1$ nevertheless require the full effective
potentials. Accordingly, the detailed high-overtone trajectories throughout the
$\epsilon=0.1$ scan are identified as results of the explicitly stated first-order dual
effective model, rather than as full-potential predictions.

\begin{figure*}[htb]
    \begin{center}
    \subfigure[~($n=1/3,a=7/\beta$)]  {\label{deltaVz-e}
    \includegraphics[width=5.6cm]{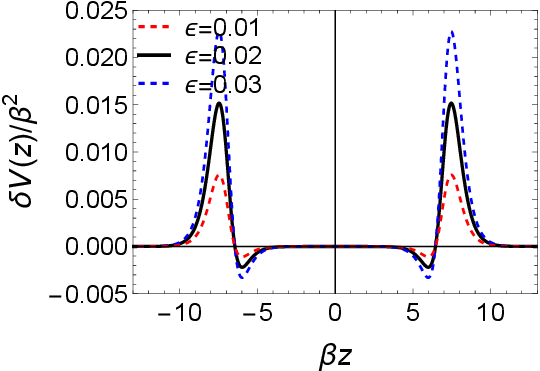}}
    \subfigure[~($n=1/3,\epsilon=0.01$)]  {\label{deltaVz-a}
    \includegraphics[width=5.6cm]{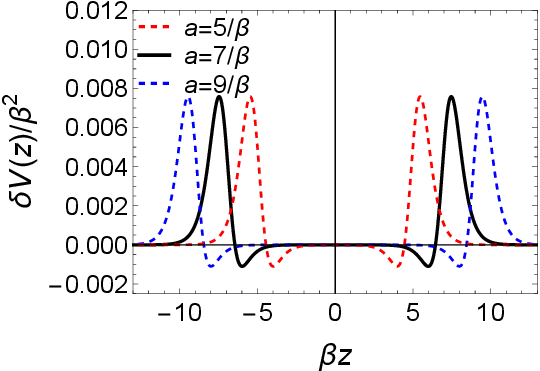}}
    \subfigure[~($\epsilon=0.01,a=7/\beta$)]  {\label{deltaVz-n}
    \includegraphics[width=5.6cm]{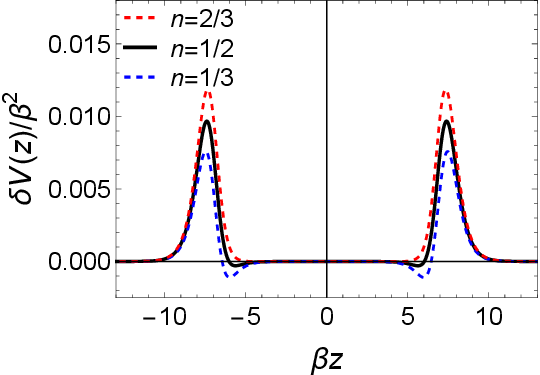}}
    \end{center}
    \caption{Specific shapes of the deformation of the effective potential~\eqref{perturbed-effective-potential}.}
    \label{figure-5}
\end{figure*}

Motivated by the effective potential deformations 
induced by the kink-type perturbations analyzed in the previous section (see Fig.~\ref{figure-2}),
we propose an explicit parametrization of the superpotential designed to reproduce these deformations.
The parametrized form reads
\begin{equation}%
    \label{parametrized-form}
    \delta W(z;\epsilon,a) = \frac{\epsilon \beta}{2}\left\{\tanh^2[\beta(z+a)]-\tanh^2[\beta(z-a)]\right\},
\end{equation}
where $\epsilon$ and $a$ control the amplitude and the position of the perturbation, respectively.
Figure~\ref{figure-5} illustrates the deformation of the effective potential 
induced by the perturbation~\eqref{parametrized-form} 
within the parametrized superpotential framework.
A direct comparison with Fig.~\ref{figure-2} shows the common features used
in the parametrization: even symmetry, two localized structures, and separate control of their
strength and position. This is a qualitative correspondence of deformation classes, not a
numerical map between the two parameter sets.

The two constructions should not, however, be interpreted as quantitatively identical. The
parameters $(\epsilon_m,a_m)$ specify the matter perturbation in
Eq.~\eqref{phi-kink-perturbation}, after which the warp factor and physical tensor potential are
reconstructed nonlinearly from the Einstein--scalar equations. By contrast, $(\epsilon,a)$ label
an analytic superpotential proxy used to explore the spectral consequences of the observed
double-barrier response. There is consequently no universal identification
$\epsilon=\epsilon_m$ or $a=a_m$.

Establishing a quantitative correspondence would require a separate case-by-case comparison of
the full potential profiles, with a specified comparison interval, norm or weight, parameter
domain, and reported residual. The present analysis does not impose such a calibration; the
QNM scans instead vary $(\epsilon,a)$ independently. Their results characterize the
Einstein--scalar-motivated double-barrier effective model and are not assigned to a unique
physical pair $(\epsilon_m,a_m)$.

We now turn to a detailed analysis of 
how this parametrized perturbation modifies the spectrum associated with the original effective potential, 
and how these spectral changes manifest in the time-domain evolution of gravitational perturbations.

\subsection{Frequency-domain} \label{Frequency-domain}

In what follows, we compute the QNM spectrum associated with
the perturbed effective potential by solving Eq.~\eqref{extra-dimensional-equation-re}
in the frequency domain with the boundary conditions 
\begin{equation}%
    \label{boundary-condition}
    \psi(z) \propto \left\{
    \begin{aligned}
        & e^{i \omega z},  &&z\rightarrow \infty, \\
        & e^{-i \omega z},  &&z\rightarrow -\infty,
    \end{aligned}
    \right.
\end{equation}
This frequency-domain reduction is exact for $p=0$ and describes the late-time regime for
$p\neq0$ when $e^{-2\alpha(t-t_0)}p^2\ll1$.
To facilitate the application of semi-analytical methods in the subsequent analysis, 
we introduce a new coordinate variable that maps 
the nontrivial hyperbolic functions appearing in the equation into polynomials. 
Accordingly, we define the change of variables $u = \tanh(\beta z)$,
under which the equation and boundary conditions take the form, respectively,
\begin{equation}%
    \left(1- u^2\right) \partial_u^2 \psi(u) - 2u \partial_u \psi(u) + \frac{\omega^2-V_{\text{re}}(u;\epsilon,a,n)}{\beta^2 \left(1- u^2\right)} \psi(u) = 0
\end{equation}
and
\begin{equation}%
    \label{boundary-condition-re}
    \psi(u) \propto \left\{
    \begin{aligned}
        & (1- u)^{-\frac{i \omega}{2\beta}},  && u\rightarrow 1, \\
        & (1+ u)^{-\frac{i \omega}{2\beta}},  && u\rightarrow -1.
    \end{aligned}
    \right.
\end{equation}
Employing the asymptotic behavior implied by the boundary conditions, 
we decompose the wave function as
\begin{equation}%
    \psi(u) = \left[(1-u) (1+u)\right]^{-\frac{i \omega}{2\beta}} \tilde{\psi}(u),
\end{equation}
which yields the equation for $\tilde{\psi}(u)$, 
\begin{equation}%
    \label{psi-equation-re}
    c_{1}(u) \partial_u^2 \tilde{\psi}(u) + c_{2}(u) \partial_u \tilde{\psi}(u) + c_{3}(u) \tilde{\psi}(u)  =0, 
\end{equation}

where $\sigma=-,+$ labels the original and dual potentials, respectively,
$V_{\mathrm{re},\sigma}^{(1)}=V_\sigma^{(1)}-9\alpha^2/4$, and
\begin{align}%
    c_{1} =& \left(1-u^2\right)^2,\\
    c_{2} =& -2 u \left(1-u^2\right) (1- \frac{i\omega}{\beta}),\\
    c_{3,\sigma}^{(1)} =&
    \left(1-u^2\right)
    \left(\frac{\omega^2}{\beta^2}+\frac{i\omega}{\beta}\right)
    -\frac{V_{\mathrm{re},\sigma}^{(1)}(u)}{\beta^2},\\
    S = & \tanh(\beta a). \nonumber
\end{align}
The reported first-order frequency-domain spectra use $c_3=c_{3,+}^{(1)}$, corresponding to the
first-order dual potential defined above. For either sign, retaining the exact quadratic term
in the deformed superpotential amounts to the replacement
\begin{equation}%
 c_{3,\sigma}^{\mathrm{full}}(u)=c_{3,\sigma}^{(1)}(u)
 -\frac{4\epsilon^2 S^2u^2(1-u^2)^2(1-S^2)^2}
 {(1-S^2u^2)^4},
 \label{full-c3}
\end{equation}
which is the expression used in the full-potential cross-check reported after
Eq.~\eqref{quadratic-ratios}.

In order to solve Eq.~\eqref{psi-equation-re}, we employ semi-analytical methods 
widely adopted in black hole QNM calculations, 
including the shooting method~\cite{Pani:2013pma} and the Bernstein spectral method~\cite{Fortuna:2020obg}.

It is worth emphasizing that our primary interest concerns 
how the perturbation modifies the QNM spectrum in the frequency domain. 
For convenience, we parameterize the perturbed quasinormal frequencies (QNFs) as
$\omega_N^{(\epsilon)} = \operatorname{Re}\left(\omega_N^{(\epsilon)}\right) + i \operatorname{Im}\left(\omega_N^{(\epsilon)}\right)$,
where $N$ labels the overtones.

For any mode branch that can be tracked continuously from $\epsilon=0$ within the restricted
family~\eqref{parametrized-form}, we quantify mode migration by
\begin{equation}%
 R_N(\epsilon,a)=
 \frac{|\omega_N^{(\epsilon,a)}-\omega_N^{(0)}|}{|\omega_N^{(0)}|},
 \qquad
 \eta_V(\epsilon,a)=
 \frac{\lVert\delta V^{(1)}(\epsilon,a)\rVert_\infty}{\lVert V_0\rVert_\infty}.
 \label{finite-family-diagnostics}
\end{equation}
The quantities $R_N$ and $R_N/\eta_V$ are finite-family response diagnostics; we do not
impose an arbitrary binary stability threshold. They are not an $\epsilon$-pseudospectrum,
because the ansatz parameter $\epsilon$ is not an operator norm. A genuine pseudospectral
statement requires a specified norm, operator domain, and outgoing boundary
realization~\cite{Jaramillo:2020tuu}.

The QNMs associated with the unperturbed effective potential ($\epsilon=0$)
have been calculated in our previous work~\cite{Jia:2024sdk}, 
with the resulting spectrum given by
\begin{equation}%
    \label{QNMs-unperturbed}
    \omega^{(\epsilon=0)} = \left(\pm\frac{3n+1}{2}-\frac{2l+1}{2}\right)i \beta, \quad l=0,1,2,\ldots.
\end{equation}

To assess the numerical reliability of the nonvanishing real parts, we combine a
cross-method comparison with method-specific convergence tests. Table~\ref{table-qnm-method-comparison}
compares the first five complex roots obtained independently with the Bernstein and shooting
methods for two representative parameter sets. All frequencies and differences are normalized
by $\beta$, and $\Delta\omega_N\equiv\omega_N^{\mathrm B}-\omega_N^{\mathrm S}$.

\begin{table}[H]
    \renewcommand\arraystretch{1.15}
    \centering
    
    \small
    \setlength{\tabcolsep}{4.5pt}
    \begin{tabular}{|c|c|c|c|c|}
        \hline
        \multirow{2}{*}{$N$} & \multirow{2}{*}{Bernstein spectral method}
        & \multirow{2}{*}{Shooting method} & \multicolumn{2}{c|}{Absolute difference} \\
        \cline{4-5}
        & & & $|\Delta\operatorname{Re}\omega_N|/\beta$ & $|\Delta\omega_N|/\beta$ \\
        \hline
        \multicolumn{5}{|c|}{$(n,\epsilon,\beta a)=(1/3,0.1,5)$} \\
        \hline
        $1$ & $0.326761-0.152508i$ & $0.326274-0.153405i$ & $4.87\times10^{-4}$ & $1.02\times10^{-3}$ \\
        $2$ & $0.527408-0.264505i$ & $0.529673-0.265976i$ & $2.27\times10^{-3}$ & $2.70\times10^{-3}$ \\
        $3$ & $0.865766-0.303535i$ & $0.866662-0.301952i$ & $8.96\times10^{-4}$ & $1.82\times10^{-3}$ \\
        $4$ & $1.147560-0.385982i$ & $1.145970-0.387304i$ & $1.59\times10^{-3}$ & $2.07\times10^{-3}$ \\
        $5$ & $1.441870-0.452909i$ & $1.451590-0.461273i$ & $9.72\times10^{-3}$ & $1.28\times10^{-2}$ \\
        \hline
        \multicolumn{5}{|c|}{$(n,\epsilon,\beta a)=(1/3,0.01,5)$} \\
        \hline
        $1$ & $0.249268-0.338966i$ & $0.242234-0.341501i$ & $7.03\times10^{-3}$ & $7.48\times10^{-3}$ \\
        $2$ & $0.493937-0.513012i$ & $0.506643-0.521727i$ & $1.27\times10^{-2}$ & $1.54\times10^{-2}$ \\
        $3$ & $0.802350-0.517096i$ & $0.811444-0.511426i$ & $9.09\times10^{-3}$ & $1.07\times10^{-2}$ \\
        $4$ & $1.093450-0.604416i$ & $1.088820-0.590693i$ & $4.63\times10^{-3}$ & $1.45\times10^{-2}$ \\
        $5$ & $1.380930-0.665379i$ & $1.375480-0.635090i$ & $5.45\times10^{-3}$ & $3.08\times10^{-2}$ \\
        \hline
    \end{tabular}
    \caption{Comparison of the first five complex QNMs obtained with the
    Bernstein spectral method and the shooting method. The parameter tuple is
    $(n,\epsilon,\beta a)$, and $\Delta\omega_N=\omega_N^{\mathrm B}-\omega_N^{\mathrm S}$.}
    \label{table-qnm-method-comparison}
\end{table}

For each numerical control parameter $\chi$, convergence is measured using the complete
complex-frequency relative difference
\begin{equation}%
    \delta\omega_N(\chi)
    =\frac{\left|\omega_N(\chi+\Delta\chi)-\omega_N(\chi)\right|}
    {\left|\omega_N(\chi)\right|},
    \qquad \chi\in\{B,\mathrm{Pre},z_{\max}\},
    \label{qnm-relative-error}
\end{equation}
as an overall measure of the stability of the complex root. This quantity does not by itself
establish whether $\operatorname{Re}\omega$ is nonzero. Because the real part is central to
the present result, Table~\ref{table-qnm-method-comparison} also reports the separate absolute
real-part discrepancy $|\Delta\operatorname{Re}\omega_N|/\beta$, which is compared directly
with the observed real parts; the total complex discrepancy $|\Delta\omega_N|/\beta$ is
retained as a conservative cross-method scale. We do not normalize the real-part difference
by $\operatorname{Re}\omega$, which would become ill-conditioned near the purely imaginary
unperturbed spectrum. Here $B$ is the Bernstein polynomial order, $\mathrm{Pre}$ is the
working-precision parameter used in the direct integration, and $z_{\max}$ is the outer
truncation position.

\begin{figure}[H]
    \centering
    \subfigure{\label{figure-qnm-convergence-B}
    \includegraphics[width=0.318\textwidth]{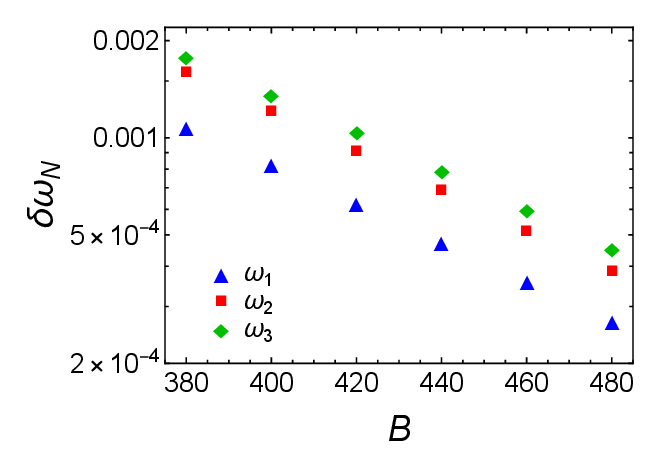}}
    \hfill
    \subfigure{\label{figure-qnm-convergence-Pre}
    \includegraphics[width=0.318\textwidth]{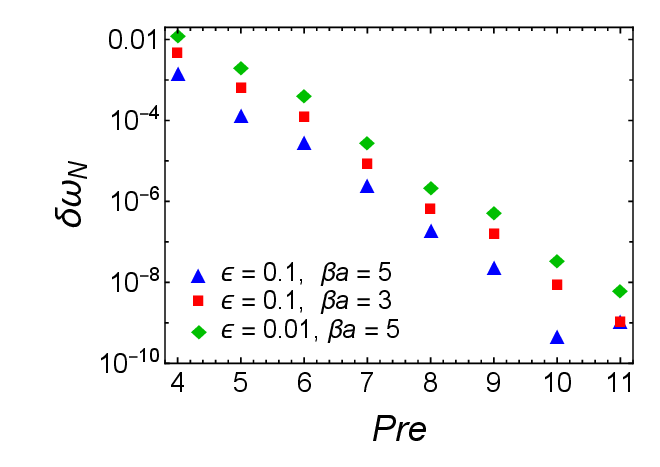}}
    \hfill
    \subfigure{\label{figure-qnm-convergence-zmax}
    \includegraphics[width=0.318\textwidth]{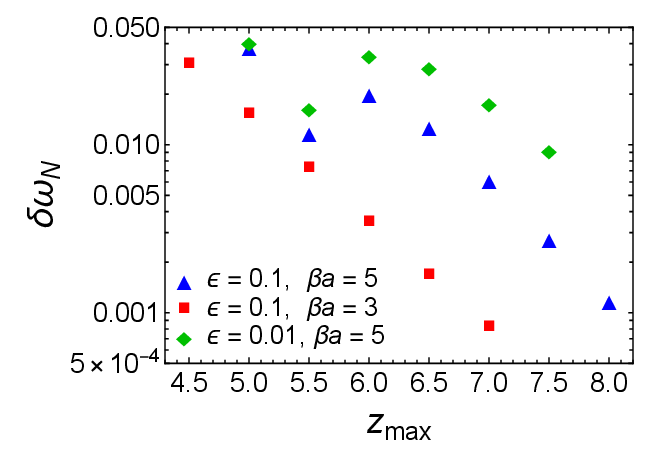}}
    \caption{Relative convergence of the complete complex QNM frequencies,
    measured using Eq.~\eqref{qnm-relative-error}. Left panel: Bernstein-order test with $\Delta B=20$ for
    the first three QNMs at $(n,\epsilon,\beta a)=(1/3,0.1,5)$. Middle panel:
    direct-integration working-precision test with $\Delta\mathrm{Pre}=1$ for
    $(\epsilon,\beta a)=(0.1,5)$, $(0.1,3)$, and $(0.01,5)$, with $n=1/3$. Right panel:
    direct-integration outer-truncation test with $\Delta z_{\max}=0.5$ for the same three
    parameter sets.}
    \label{figure-qnm-convergence}
\end{figure}

The three panels isolate complementary numerical error sources. The Bernstein-order panel
monitors the first three QNMs while the polynomial order is increased in steps of $\Delta B=20$.
The working-precision and outer-truncation panels vary the shooting parameters in steps of
$\Delta\mathrm{Pre}=1$ and $\Delta z_{\max}=0.5$, respectively, for three representative
parameter sets. In the final plotted intervals $B=480\mathbin{\to}500$ and
$\mathrm{Pre}=11\mathbin{\to}12$, the corresponding relative differences are between
$2.7\times10^{-4}$ and $4.5\times10^{-4}$ for the first three Bernstein roots, and between
$1.0\times10^{-9}$ and $6.6\times10^{-9}$ for the working-precision test. For the three
outer-truncation cases, the final intervals are respectively
$z_{\max}=8.0\mathbin{\to}8.5$, $7.0\mathbin{\to}7.5$, and
$7.5\mathbin{\to}8.0$; the corresponding differences range from
$7.0\times10^{-4}$ to $9.3\times10^{-3}$. Across the first five modes in
Table~\ref{table-qnm-method-comparison}, the real-part discrepancy
$|\Delta\operatorname{Re}\omega_N|/\beta$ ranges from $4.87\times10^{-4}$ to
$9.72\times10^{-3}$ for $\epsilon=0.1$, while $|\Delta\omega_N|/\beta$ ranges from
$1.02\times10^{-3}$ to $1.28\times10^{-2}$. For $\epsilon=0.01$, the corresponding
ranges are $4.63\times10^{-3}$--$1.27\times10^{-2}$ and
$7.48\times10^{-3}$--$3.08\times10^{-2}$, respectively. The observed nonzero real parts
exceed the conservative total discrepancies by factors ranging from approximately
$3.2\times10^1$ to $5.5\times10^2$. Thus, the cross-method scale remains well below the
observed real parts for all ten tabulated roots. The agreement of the two independent methods,
together with their separate convergence behavior, confirms that the emergence of a
nonvanishing real part is not a numerical truncation or root-finding artifact.

\begin{figure*}[htb]
    \begin{center}
    \subfigure{\label{figure-6-1}
    \includegraphics[width=7.6cm]{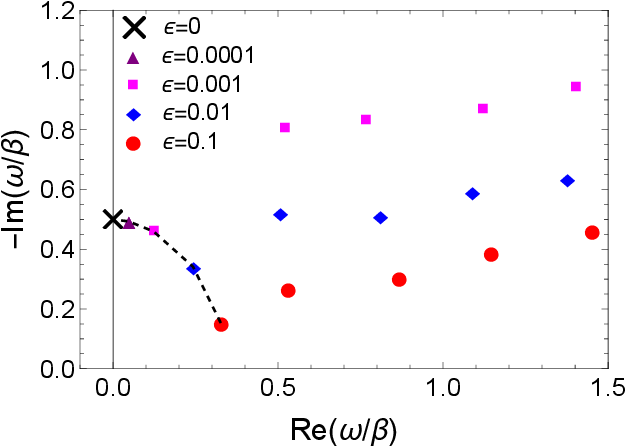}}
    \subfigure{\label{figure-6-2}
    \includegraphics[width=7.6cm]{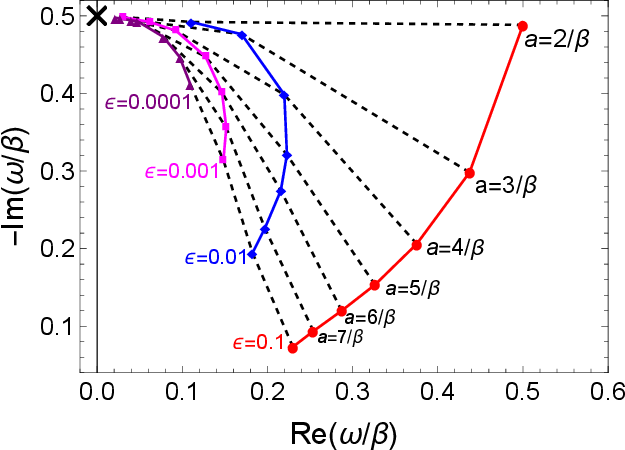}}
    \end{center}
    \caption{Left panel: Variation of the QNM spectrum
    induced by the perturbation for $n=1/3$ and $a=5/\beta$.
    The black cross denotes the first purely imaginary QNM of the
    unperturbed effective potential, while the black dashed line connects
    the selected first-QNM markers obtained in the original scan for different
    perturbation amplitudes.
    Right panel: Dependence of the selected first-QNM marker on the perturbation
    location $a$ for different perturbation amplitudes. Each black dashed line
    corresponds to a fixed value of $a$.
    The two original graphical scans are retained unchanged for direct
    comparison with the submitted manuscript. The numerical validation of the selected
    complex roots is reported in Table~\ref{table-qnm-method-comparison} and
    Fig.~\ref{figure-qnm-convergence}.}
    \label{figure-6}
\end{figure*}

We now turn to the impact of perturbations on the QNM spectrum, 
with particular emphasis on the role of the perturbation amplitude $\epsilon$.
The left panel of Fig.~\ref{figure-6} illustrates the distribution of QNMs 
for different values of $\epsilon$, with $n=1/3$ and $a=5/\beta$. 
The black cross denotes the least-damped QNM $\omega_1^{(\epsilon=0)}$
of the unperturbed effective potential, which is purely imaginary. 
  The markers in this panel reproduce the original graphical scan.
  Because more than one pole may coexist
 at a given parameter value, the graphical selection is not used to assert a unique
 least-damped ordering for every weak-deformation point.
Moreover, as the perturbation amplitude increases, the QNMs deviate progressively farther
from their original positions in the complex frequency plane.

Furthermore, for sufficiently strong perturbations, a sequence of higher-overtone modes 
with approximately uniform spacing appears in the spectrum. 
This behavior can be ascribed to the double-barrier-like structure of the effective potential 
generated by the parametrized perturbation in Eq.~\eqref{parametrized-form} (see also Fig.~\ref{figure-5}). 
 The black dashed line in the left panel of Fig.~\ref{figure-6} is the
 original visual guide through selected first-QNM markers. Because the ordering of the
 purely imaginary and complex branches can change, it should not be interpreted as a smooth
 continuation of one rigorously tracked analytic eigenvalue branch.

 The right panel qualitatively illustrates the migration of the selected
 complex root with $a$. The plotted loci are descriptive scan trajectories, not a proof
 that the selected marker remains the least-damped pole for every parameter value and not
 a universal circular or exponential law.

\begin{figure*}[htb]
    \begin{center}
    \subfigure{\label{figure-7-1}
    \includegraphics[width=7.6cm]{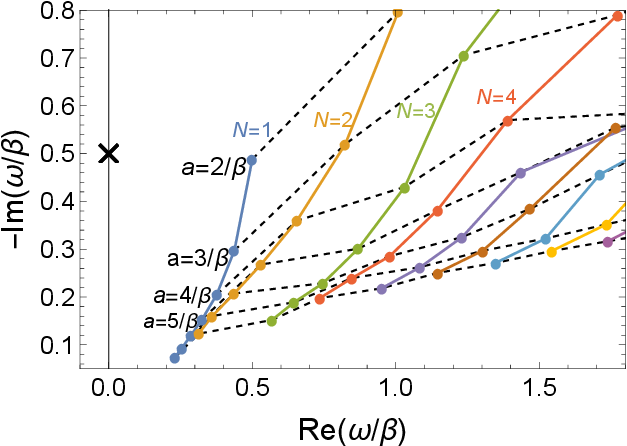}}
    \subfigure{\label{figure-7-2}
    \includegraphics[width=7.6cm]{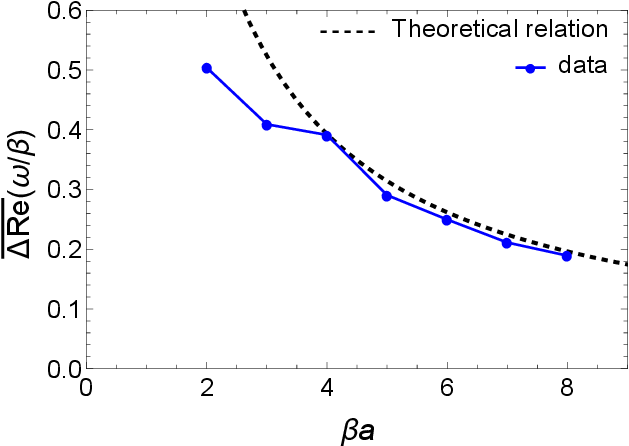}}
    \end{center}
    \caption{Left panel: Dependence of the QNMs on the perturbation location $a$
    for a fixed perturbation amplitude $\epsilon=0.1$, with $n=1/3$.
    Solid curves with different colors trace the evolution of different overtone modes
    $\omega_N^{(\epsilon=0.1)}$ as functions of $a$, while the black dashed lines
    indicate results corresponding to the same value of $a$.
    Right panel: Comparison between the average spacing of the real parts of the QNFs
    for different values of $a$ and the corresponding theoretical prediction.
    The original two-panel graphical scan is retained unchanged.
    The comparison is used only as qualitative evidence for the large-separation
    cavity trend.}
    \label{figure-7}
\end{figure*}

We next investigate the influence of the perturbation location parameter $a$
on the QNM spectrum. Focusing on the case $\epsilon=0.1$ and $n=1/3$,
we analyze the behavior of higher-overtone modes. 
As shown in the left panel of Fig.~\ref{figure-7}, 
increasing the distance of the perturbation from the brane results in 
an increasingly dense distribution of the real parts of the high-overtone frequencies, 
while their imaginary parts decrease in magnitude, indicating a reduced damping rate.

In particular, the first ($N=1$) and second ($N=2$) QNMs approach each other 
as the parameter $a$ increases. 
This behavior may be related to the emergence of exceptional points 
discussed in recent studies of QNM resonances in black hole systems~\cite{Motohashi:2024fwt,Cavalcante:2024swt,Yang:2025dbn,PanossoMacedo:2025xnf,Wu:2025wbp,Cheng:2026gxu}.
The proximity of two frequencies alone is not proof of an exceptional point;
that identification would require eigenfunction coalescence or a square-root branch structure,
which we do not test here.
Moreover, the overall evolution of the spectrum with respect to $a$ 
exhibits large parameter-induced spectral migration: as the perturbation is placed farther away, 
the spectrum deviates increasingly from its unperturbed configuration. 
This trend is consistent with our previous findings on the stability of 
QNM spectra in flat brane backgrounds~\cite{Jia:2025saq}.

The underlying physical mechanism can be ascribed to the double-barrier-like 
structure of the effective potential induced by the perturbation. 
As the separation between the two barriers increases, 
the potential well can accommodate a larger number of quasi-bound states, 
giving rise to the additional QNMs. 
To support this interpretation, we further examine the spacing between 
the real parts of adjacent QNFs at fixed values of $a$.
We observe that the spacing approximately follows the equal-spacing relation characteristic
of resonant states in a double-barrier system
\cite{Cardoso:2016rao,Mark:2017dnq,Tan:2024qij,Jia:2025saq},
\begin{equation}%
    \Delta \operatorname{Re}(\omega) \sim \frac{\pi}{2a},
    \label{cavity-spacing}
\end{equation}
as illustrated in the right panel of Fig.~\ref{figure-7}.
 The closer agreement at larger separation supports the leaky-cavity
 interpretation, while the remaining mismatch is naturally attributable to finite-barrier
 reflectivity and phase-shift corrections. Because the original scan was not produced from
 one uniform cutoff-convergence grid, we do not assign its plotted spacing residuals the status
 of numerical error bars. Equation~\eqref{cavity-spacing} is a qualitative large-separation
 trend, not an exact law.

 In summary, the independently cutoff-tested roots establish complex-frequency pairs within
 the stated nonzero deformations; they do not imply that the least-damped QNM is complex at
 every sampled point. At $(n,\epsilon,\beta a)=(1/3,0.001,5)$, the least-damped pole remains purely imaginary,
while the leading complex pair lies slightly deeper in the spectrum. The separated reflective
structures are the relevant condition for the approximately equally spaced cavity spectrum in
the present mechanism. Other localized deformations can also produce complex QNMs with
nonzero real parts, but their spectral organization need not follow
Eq.~\eqref{cavity-spacing} and must be determined from the corresponding potential.

We find that the appearance of the real part is highly sensitive to 
both the perturbation strength $\epsilon$ and its location $a$. 

The branch-resolved results instead exhibit a competition between a surviving
purely imaginary branch and the complex cavity branch. Their crossing in damping rate explains
why the identity of the least-damped pole can change with $\epsilon$ and $a$; this identification
is based on separately resolved roots rather than solely on the graphical trajectories in
Fig.~\ref{figure-6}.

Moreover, a systematic analysis of higher-overtone modes reveals that 
the spacing between the real parts of adjacent QNFs approximately 
obeys the theoretical relation expected for a double-barrier potential. 
This observation provides further support for the interpretation that 
the perturbation-induced double-barrier-like structure of the effective potential 
plays a key role in shaping the QNM spectrum.

This interpretation also clarifies the relation to black-hole spectral instability.
The large motion of QNMs under a small change of the potential reflects the general sensitivity
of a non-self-adjoint and generally non-normal outgoing problem, as quantified through its
resolvent and pseudospectrum~\cite{Jaramillo:2020tuu}. The particular direction of
motion, the approximately equal spacing, and the appearance of long-lived branches in the
present model are more specific: they result from the finite cavity formed by the
braneworld-induced double-barrier structure. 
We therefore regard the non-self-adjoint sensitivity as the
general mathematical setting and the double-barrier geometry as the physical realization in
this dS braneworld. A full pseudospectrum calculation would require a specified operator norm,
domain, and compactification and is left for future work.

In the following, we turn to a time-domain analysis to investigate 
how these QNMs with nonzero real parts manifest 
in the corresponding gravitational-wave signals.

\subsection{Time-domain} \label{Time-domain}

The above frequency-domain analysis demonstrates that the double-barrier
perturbations considered here generate nonvanishing real parts
in the QNM spectrum and may significantly alter the distribution of overtones, 
especially when the perturbation is located far from the brane. 
While these spectral features demonstrate substantial eigenfrequency migration
within the sampled family, their direct observational relevance ultimately depends on 
how they are imprinted on the time-domain gravitational-wave signal. 
In particular, it is essential to understand whether the emergence of real-frequency components 
    produces resolvable changes in the normalized model waveform, 
    or whether such effects are suppressed during the dynamical evolution. 
Motivated by these considerations, we now proceed to a time-domain analysis to examine 
how the perturbation-induced modifications of the QNM spectrum are manifested in
the temporal evolution of gravitational perturbations.

Specifically, the time-domain analysis requires solving Eq.~\eqref{psi-equation}
in conjunction with Eq.~\eqref{hij-assumption}, from which the evolved waveform $\Phi(t,z)$ is extracted.
This is the model waveform analyzed below, not detector-calibrated strain.

In addition to evolving the original equation, we adopt the dual-equation diagnostic of
Ref.~\cite{Jia:2024sdk} to suppress the zero-mode contribution at the retained perturbative order
and display the QNM-dominated response more transparently.

For computational convenience, we rewrite the evolution equations in light-cone coordinates,
$du=dt-dz$, $dv=dt+dz$,
which leads to the explicit form
\begin{align}%
    \label{uv-wave-equation-origin}
    \text{original:} \quad
    &\left[4 \frac{\partial^{2}}{\partial_{u}\partial_{v}} +V(u,v) + e^{-2 \alpha t} p^2 - \frac{9}{4} \alpha ^2  \right] \Psi (u,v) =0, \\
    \label{uv-wave-equation-dual}
    \text{dual:} \quad
    &\left[4 \frac{\partial^{2}}{\partial_{u}\partial_{v}} +V_{\text{dual}}(u,v) + e^{-2 \alpha t} p^2 - \frac{9}{4} \alpha ^2  \right] \hat{\Psi} (u,v) =0,
\end{align}

In the calculations shown below, $V=V_-^{(1)}$ and
$V_{\mathrm{dual}}=V_+^{(1)}$, as defined in Eq.~\eqref{first-order-partners}. Accordingly,
the dual first-order evolution is used only as a perturbative diagnostic. Numerically it
strongly suppresses the zero-mode-dominated late-time component over the simulated interval;
we do not claim that $V_+^{(1)}$ is exactly isospectral to $V_-^{(1)}$ or that the truncated
operator removes an exact zero eigenstate. Throughout the time-domain discussion, ``zero
mode'' denotes this zero-mode sector at the retained order. The full original--dual pair has
an exact intertwining relation and exact zero-mode properties.

It is worth emphasizing that the original-potential evolution within the
effective model is governed by Eq.~\eqref{psi-equation} rather than its dual counterpart.
Our primary focus is how perturbation-induced modifications of the QNM spectrum are encoded
in the model waveform, not the construction of detector-calibrated strain. From this
perspective, the dual equation is used only as a diagnostic tool.

For simplicity, we choose either odd or even Gaussian wave packets as initial data 
and perform a numerical evolution for Eqs.~\eqref{uv-wave-equation-origin} and~\eqref{uv-wave-equation-dual}. 
Specifically, the initial conditions are taken to be
\begin{align}%
    \label{odd-initial-data}
    \text{odd:} \quad &\Psi_{\text{in}}(u,0)=\sin\left(\beta u\right) e^{-\frac{\beta^2 u^2}{2}}, \quad \Psi_{\text{in}}(0,v)=\sin\left(-\beta v\right) e^{-\frac{\beta^2 v^2}{2}},\\
    \label{even-initial-data}
    \text{even:} \quad &\hat{\Psi}_{\text{in}}(u,0)=\cos\left(\beta u\right) e^{-\frac{\beta^2 u^2}{2}}, \quad \hat{\Psi}_{\text{in}}(0,v)=\cos\left(-\beta v\right) e^{-\frac{\beta^2 v^2}{2}}.
\end{align}
Furthermore, the term $e^{-2 \alpha t} p^2$ in the evolution equations effectively produces 
an overall upward shift of the effective potential, but this contribution gradually weakens with time.
As pointed out in Ref.~\cite{Jia:2024sdk}, this factor mainly affects the early-time ringdown signal, 
while its influence becomes negligible at late times.
Motivated by this observation, we analyze the time-domain evolution separately 
for the cases $p=0$ and $p\neq 0$. 

\vspace{8pt}
\textbf{(i) \textit{The case of $p=0$.}}
\vspace{5pt}

\begin{figure*}[htb]
    \begin{center}
    \subfigure{\label{timedomain-origin-varye}
    \includegraphics[width=8.0cm]{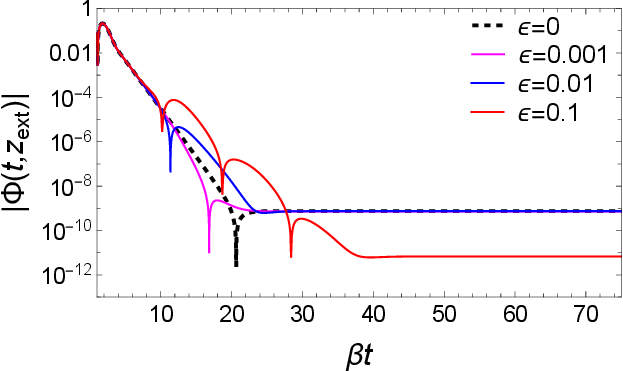}}
    \subfigure{\label{timedomain-origin-varya}
    \includegraphics[width=8.0cm]{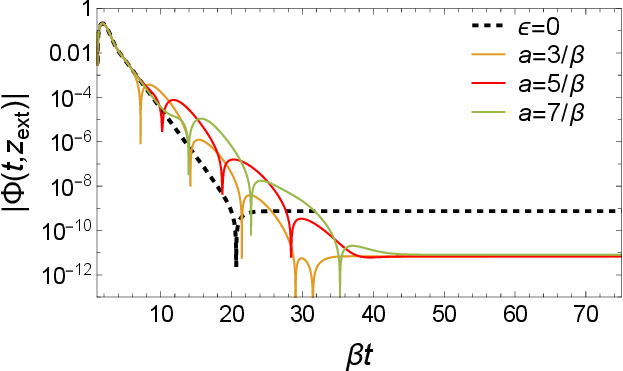}}
    \end{center}
    \caption{Time-domain evolutions for odd initial data~\eqref{odd-initial-data}
    governed by Eq.~\eqref{uv-wave-equation-origin} for $n=1/3$. 
    The waveform is extracted at $\beta z_{ext}=1$. 
    Left panel: time-domain waveforms for different values of $\epsilon$ with fixed $a=5/\beta$. 
    The black dashed curve denotes the evolution in the unperturbed case ($\epsilon=0$), 
    while the colored solid curves correspond to different values of $\epsilon$. 
    Right panel: time-domain waveforms for different values of $a$ with fixed $\epsilon=0.1$.}
    \label{figure-8}
\end{figure*}

Figure~\ref{figure-8} presents the time-domain evolution of an odd Gaussian wave packet 
governed by the evolution equation~\eqref{uv-wave-equation-origin}, 
with the extraction point chosen at $\beta z_{ext}=1$.
The black dashed curve denotes the waveform in the unperturbed background, 
while the colored solid curves distinguish the evolutions corresponding to different perturbation parameters.
It is worth emphasizing that we initially considered even initial wave packets.
However, in that case no quasinormal ringing was resolved, and the waveform
remained dominated by the zero-mode sector over the simulated interval. Adding perturbations
did not qualitatively alter this behavior.
This feature was also observed in our previous work~\cite{Jia:2024pdk,Jia:2024sdk,Jia:2025saq}. 
The underlying reason, within the retained first-order description, is that the
even zero-mode sector remains dominant whereas the first QNM is odd. Consequently, even initial
data preferentially excites the former and overwhelms the QNM signal.
To isolate the QNM contribution, typically dominated by the first QNM, 
we therefore adopt the odd initial wave packets.

From the left panel of Fig.~\ref{figure-8}, 
one can see that the unperturbed waveform (black dashed curve) is essentially composed of two stages. 
The early-time behavior exhibits a purely decaying, nonoscillatory signal, 
which is dominated by the first QNM with a purely imaginary frequency. 
At late times, the amplitude approaches an almost constant value, reflecting the dominance of
the zero-mode sector over the simulated interval.
In contrast, once perturbations are introduced, the waveform can be clearly divided into three stages. 
In addition to the early and late-time behaviors, an intermediate stage characterized by damped oscillations emerges. 
As will be demonstrated in the next section, this intermediate regime carries
the leading complex-frequency QNM component of the perturbed spectrum.
Furthermore, the oscillatory signal in this intermediate stage becomes less pronounced 
as the perturbation strength decreases.
In particular, for the blue and purple curves, the oscillation frequency (i.e., the real part of the mode) is barely distinguishable, 
while the decay rate (the imaginary part) can be approximately extracted. 
By contrast, the case with $\epsilon=0.1$ exhibits a much more pronounced oscillatory decay in the intermediate stage, 
allowing for a clearer identification of the QNM signal.
For all sampled perturbation parameters, the QNM-dominated signal is eventually
overwhelmed by the zero-mode sector within the simulated time interval.

The right panel of Fig.~\ref{figure-8} illustrates the influence of perturbations 
located at different positions on the time-domain waveform. 
As can be seen, when the perturbation is closer to the brane position ($z=0$),
the early-time stage dominated by the purely imaginary first QNM becomes shorter,
and the intermediate stage carrying the perturbation-induced signal appears earlier. 
Conversely, as the perturbation is placed farther away from the brane, 
the onset of the intermediate oscillatory stage is correspondingly delayed.

\begin{figure*}[htb]
    \begin{center}
    \subfigure{\label{timedomain-dual-varye}
    \includegraphics[width=8.0cm]{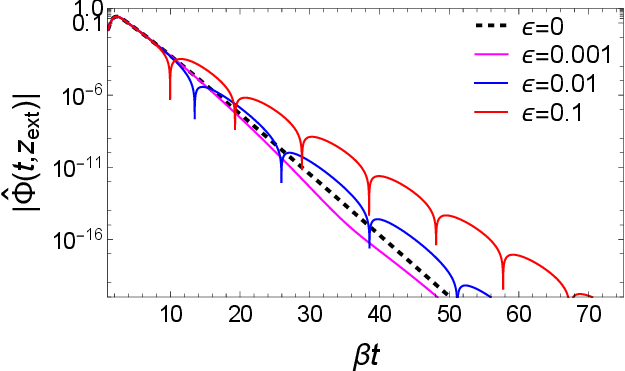}}
    \subfigure{\label{timedomain-dual-varya}
    \includegraphics[width=8.0cm]{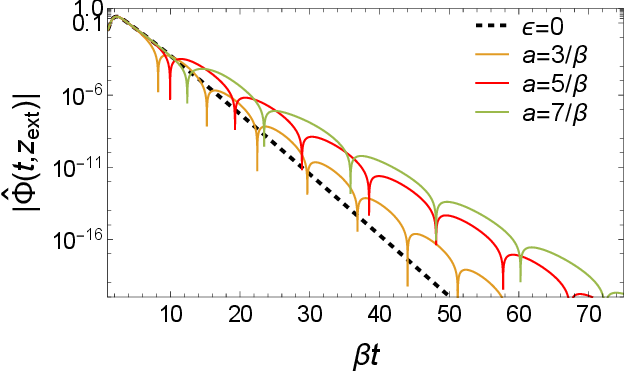}}
    \end{center}
    \caption{Time-domain evolutions for even initial data~\eqref{even-initial-data}
    governed by Eq.~\eqref{uv-wave-equation-dual} for $n=1/3$. 
    The waveform is extracted at $\beta z_{ext}=1$. 
    Left panel: waveforms for different values of $\epsilon$ with $a=5/\beta$. 
    Right panel: waveforms for different values of $a$ with $\epsilon=0.1$.}
    \label{figure-9}
\end{figure*}

Since the first-order dual evolution suppresses the zero-mode contribution at
the retained order, it provides a more transparent diagnostic of the QNM-governed waveform.
We therefore solve Eq.~\eqref{uv-wave-equation-dual} with even initial data, 
and present the resulting time-domain waveforms in Fig.~\ref{figure-9}. 
The extraction point is again chosen at $\beta z_{ext}=1$, close to the brane.

From the left panel of Fig.~\ref{figure-9}, 
one can observe that the perturbed waveform can be roughly divided into two stages. 
The early-time behavior is dominated by the unperturbed first QNM with a purely imaginary frequency, 
while the late-time evolution is governed by the leading complex-frequency
QNM component of the perturbed spectrum. 
The duration of the early-time signal is relatively short and 
depends sensitively on the location of the perturbation. 
Notably, no power-law tail, commonly observed in the time-domain evolution on flat brane backgrounds, 
appears in the present setup. In Ref.~\cite{Jia:2024sdk}, we have found that, in a de Sitter brane background, 
the late-time evolution is entirely controlled either by the zero mode or by the QNMs.
The right panel of Fig.~\ref{figure-9} illustrates the impact of the parameter $a$ on the waveform. 
The corresponding fitting results are presented in the next section. 
By comparing the results shown in Figs.~\ref{figure-8} and~\ref{figure-9}, 
one can clearly see that the difference between the evolutions governed by the two equations 
is the relative prominence of the late-time zero-mode-sector contribution:
it is strongly suppressed, rather than exactly absent, in the first-order dual evolution over
the simulated interval.
On the other hand, in Fig.~\ref{figure-8}, the QNM information induced by the perturbation 
is carried by the intermediate stage of the waveform, 
whose short duration makes an accurate extraction rather challenging. 
The evolution obtained from the dual equation, as shown in Fig.~\ref{figure-9},
mitigates this drawback and allows a clearer identification of
the perturbed QNM spectrum.

\vspace{8pt}
\textbf{(ii) \textit{The case of $p\neq 0$.}}
\vspace{5pt}

\begin{figure*}[htb]
    \begin{center}
    \subfigure{
    \includegraphics[width=8.0cm]{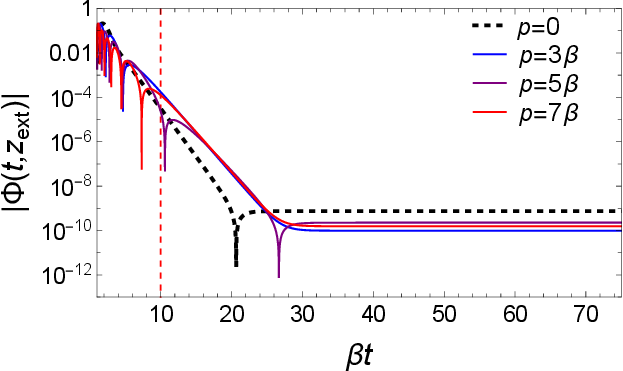}}
    \subfigure{
    \includegraphics[width=8.0cm]{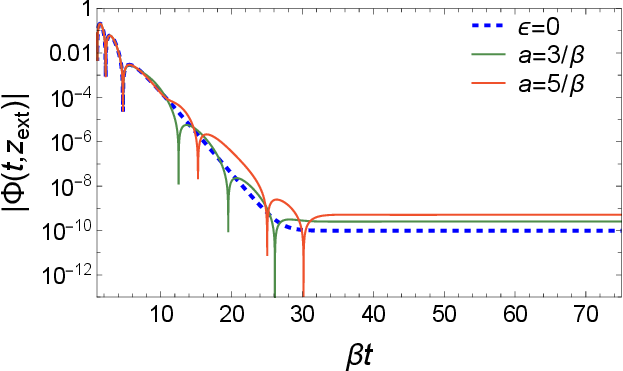}}
    \end{center}
    \caption{Time-domain evolutions for odd initial data~\eqref{odd-initial-data}
    governed by Eq.~\eqref{uv-wave-equation-origin} for $n=1/3$. 
    The waveform is extracted at $\beta z_{ext}=1$. 
    Left panel: waveforms for various values of $p$ with $\epsilon=0$. 
    Right panel: waveforms for various values of $a$ with $\epsilon=0.1$ and $p=3\beta$.}
    \label{figure-10}
\end{figure*}

To investigate the influence of the three-dimensional momentum $p$
on the time-domain waveforms, we first calculate the evolution in the absence of perturbations 
for several representative values of $p$, as shown in the left panel of Fig.~\ref{figure-10}. 
The black dashed curve corresponds to the case $p=0$,
while the colored solid curves represent different nonzero values of $p$. 

As illustrated in the left panel of Fig.~\ref{figure-10}, 
the effect of the momentum parameter $p$ is mainly confined to the early-time stage of the waveform, 
approximately before the red vertical line. During this stage, the signal exhibits oscillatory behavior 
whose frequency gradually decreases with time. 
In the intermediate stage, the waveform transitions to a purely decaying phase, 
which is associated with the first QNM possessing a purely imaginary frequency. 
Finally, the late-time evolution is dominated by the zero mode.

The early-time behavior can be understood as follows. The time-dependent positive term
$e^{-2\alpha t}p^2$ in Eq.~\eqref{uv-wave-equation-origin} raises the instantaneous effective
frequency at early times and produces the chirp-like oscillation in the model waveform. As
this shift decays exponentially, the oscillation frequency decreases, and the waveform
returns to zero-mode-sector dominance over the simulated interval.

Motivated by the significant impact of $p$ on the early-time signal, 
we further compute the time-domain evolution in the presence of perturbations, 
as shown in the right panel of Fig.~\ref{figure-10}. 
As expected, the early-time waveform is predominantly determined by the momentum parameter $p$,
with the effect of the perturbation being almost negligible. 
In the intermediate stage, as the influence of $p$ diminishes with time, 
the effect of the perturbation becomes apparent, and the waveform is mainly governed by
the leading complex-frequency QNM component. 
At late times, the waveform is once again dominated by the zero mode.

To summarize, in the previous section we investigated the effects of perturbations 
on the QNM spectrum and found that, within the double-barrier family studied,
a sufficiently strong deformation produces a large number of modes with nonvanishing real parts. 
This behavior is in sharp contrast with the unperturbed case, 
where the spectrum consists solely of a series of purely imaginary QNMs, 
leading to qualitatively different spectral properties.

In the present section, we have shown that the perturbation-induced modifications 
of the spectrum are also reflected in the time-domain signals. 
In particular, their imprint mainly manifests itself during the intermediate stage of the waveform, 
where an oscillatory and decaying signal emerges. When the perturbation strength is sufficiently small, 
however, this stage becomes too weak to clearly exhibit the effect of the perturbation.
We further examined the influence of the three-dimensional momentum $p$ on the waveform. 
Consistent with our previous results~\cite{Jia:2024sdk}, the parameter $p$ affects only the early-time behavior of the signal, 
which is characterized by a quasinormal ringing phase with a gradually decreasing oscillation frequency, 
while its impact on the intermediate stage is rather limited.

Moreover, since the perturbation considered here resembles a double-barrier potential, 
one may expect the appearance of echo signals in the waveform. 
Nevertheless, no such echoes are observed. This can be attributed to two main reasons:
the perturbation is not located sufficiently far from the brane, and any emerging echo signal
is overwhelmed by the constant-amplitude contribution associated with the zero mode.
Finally, regarding the late-time behavior of the waveform, for
the original-potential evolution within the effective model
governed by Eq.~\eqref{uv-wave-equation-origin}, the signal is dominated by the
zero-mode sector over the simulated interval. 
In contrast, for the dual equation~\eqref{uv-wave-equation-dual}, 
the late-time waveform is mainly determined by the leading complex-frequency
QNM component over the simulated time interval at the retained order, 
as clearly illustrated by the comparison between Figs.~\ref{figure-8} and~\ref{figure-9}.

\subsection{Comparative analysis and discussion of results} \label{analysis and discussion}

In the following, we perform a fitting analysis of the time-domain waveforms 
obtained in the previous section. By adopting the fitting ansatz 
\begin{equation}%
    \label{fit-timedomain}
    \Phi(t) \sim e^{-\beta t/2} e^{\operatorname{Im}(\omega)t} 
    \sin\left(\operatorname{Re}(\omega)t-\theta \right),
\end{equation}
we extract the mode frequencies encoded in the model waveform and
determine the QNFs that can be reliably resolved from the time-domain data. 
The fitted frequencies are then compared with the QNMs independently 
calculated in the frequency domain.
It should be noted that the fitting results depend sensitively 
on the choice of the fitting window.

\begin{table}[!htb]
    \renewcommand\arraystretch{1.0}
    \centering

    \begin{tabular}{|c|c|c|c|}
    \hline
    \multirow{2}{*}{} & \multicolumn{2}{c|}{~~~~~~~Time-domain~~~~~~~} &
    \multirow{2}{*}{\shortstack{Direct-integration frequency-domain\\$V_+^{(1)}$ shooting}} \\
    \cline{2-3}
    & original Eq.~\eqref{uv-wave-equation-origin} & dual Eq.~\eqref{uv-wave-equation-dual} & \\
    \hline
    ~~$\epsilon$~~  &~$\text{Re}(\omega/\beta) $~~~~$\text{Im}(\omega/\beta) $~&~$\text{Re}(\omega/\beta) $~~~~$\text{Im}(\omega/\beta) $~&~$\text{Re}(\omega/\beta) $~~~~$\text{Im}(\omega/\beta) $~  \\
    \hline
    ~~$0$~~ &~  0.000487~~~~-0.508738 ~&~   0.001466~~~~-0.499982 ~& 0~~~~-0.5  \\
    \hline
    ~~$0.001$~~ &~  0.0247498~~~~-0.368233 ~&~   0.0967193~~~~-0.542292 ~&
    \shortstack{$0\quad -0.49988$ (least)\\
    $\pm0.159502\quad -0.513397$ (complex)}  \\
    \hline
    ~~$0.01$~~ &~  0.191071~~~~-0.411263 ~&~   0.252136~~~~-0.336975 ~&
    0.249204~~~~-0.339307  \\
    \hline
    ~~$0.1$~~ &~   0.321139~~~~-0.131438 ~&~   0.326107~~~~-0.153556 ~&
    0.326814~~~~-0.152781  \\
    \hline
    \end{tabular}

    \caption{Comparison between the QNF values extracted from
    the time-domain oscillatory windows and the relevant directly integrated
    low-lying frequency-domain roots with and without the perturbation for $p=0$.
    The comparison with the original $V_-^{(1)}$ evolution is perturbative;
    the dual time-domain column and the frequency-domain column use $V_+^{(1)}$.
    At $\epsilon=0.001$, the time-domain oscillatory component is compared with the complex
    pair, not with the less-damped purely imaginary pole. The tabulated frequency-domain values were obtained
    by direct integration and are independent of the unchanged graphical markers retained in
    Fig.~\ref{figure-6}.}
    \label{Table-1}
\end{table}

We first investigate the fitting results of the time-domain waveforms 
in the case $p=0$, as summarized in Table~\ref{Table-1}.
For the unperturbed waveform, it is evident that the early-time evolution is indeed 
dominated by the first purely imaginary QNM, resulting in a purely decaying, non-oscillatory signal.
After introducing perturbations, the results obtained from the evolution equation~\eqref{uv-wave-equation-origin} 
show that, due to the presence of the zero-mode contribution, 
the duration of the intermediate stage is relatively short. In particular, for small perturbation amplitudes, 
the fitted QNFs exhibit significant deviations from the frequency-domain results, 
and in some cases a complete oscillation cycle is not even formed (see the left panel of Fig.~\ref{figure-8}). 
By contrast, the waveform corresponding to $\epsilon=0.1$ shows much better agreement 
with the frequency-domain spectrum.
For the evolution governed by the dual equation~\eqref{uv-wave-equation-dual}, 
the zero-mode-dominated component is strongly suppressed over the simulated
interval, within the accuracy of the first-order scheme. As a result, the intermediate and late-time stages 
of the waveform preferentially expose the leading complex-frequency component
of the perturbed spectrum. 
The corresponding fitting results are therefore in better agreement with those obtained 
in the frequency domain. Nevertheless, when the perturbation amplitude is sufficiently small, 
accurately extracting the QNFs remains challenging.
In particular, at $\epsilon=0.001$ this finite-window oscillatory component
is not the least-damped pole; the latter remains approximately purely imaginary.

\begin{figure*}[htb]
    \begin{center}
    \includegraphics[width=8.0cm]{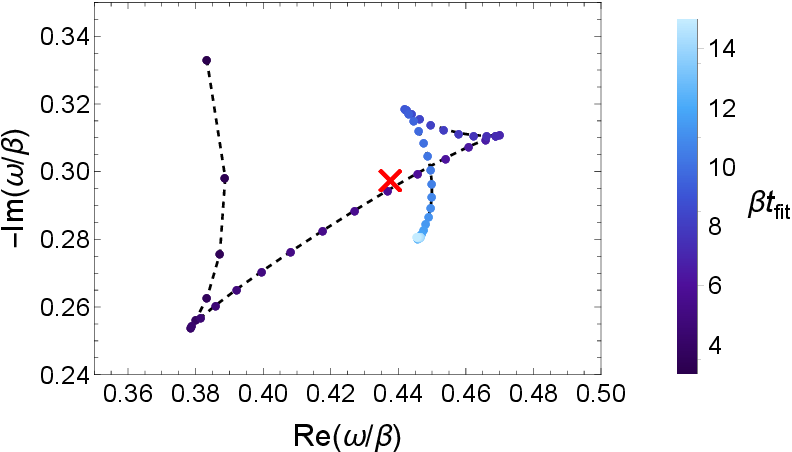}
    \end{center}
    \caption{Comparison between the time-domain evolution and 
    the frequency-domain results for the case $p=3\beta$. 
    The fitted values are extracted from the green solid curve shown 
    in the right panel of Fig.~\ref{figure-10}. 
    The red cross denotes the theoretical QNF calculated in the frequency domain. 
    The colored points correspond to fits performed over different fitting intervals 
    $(t_{\text{fit}},t_{\text{end}}=25/\beta)$.}
    \label{figure-11}
\end{figure*}

It is worth emphasizing that the early-time signal associated with the unperturbed spectrum,
particularly with its first QNM, can generally be fitted with relatively high precision. 
However, the above discussion is restricted to the case $p=0$. 
As shown in Fig.~\ref{figure-10}, the three-dimensional momentum $p$ significantly affects 
the early-time waveform, effectively obscuring the signature of the unperturbed spectrum, 
while its influence on the intermediate-time regime is much weaker. 
To confirm this observation, we perform fits over varying fitting intervals for the data 
corresponding to the green solid curve in the right panel of Fig.~\ref{figure-10}; 
the results are presented in Fig.~\ref{figure-11}.

In Fig.~\ref{figure-11}, the data points ranging from dark blue to light blue 
correspond to fitting intervals that progressively shift toward the intermediate stage of 
the waveform evolution. The fitting end time is fixed at the onset of the zero-mode contribution, 
chosen as $t_{\text{end}}=25/\beta$.
As shown in Fig.~\ref{figure-11}, when the fitting window moves toward the intermediate-time regime,
the fitted frequencies in the complex plane gradually approach the QNF obtained from 
the frequency-domain calculation (indicated by the red cross). 
It is worth noting that this convergence does not proceed monotonically along a straight trajectory. 
Instead, the fitted frequencies trace a bent path in the complex frequency plane.
This behavior can be attributed to two main factors. First, the presence of the momentum parameter $p$ 
introduces additional interference in the early-time signal. 
Second, as the fitting interval becomes narrower, the amount of available data is reduced, 
which degrades the quality of the fit and reflects the intrinsic 
uncertainties and errors associated with the fitting procedure.
Nevertheless, the overall trend (namely, the gradual convergence toward the frequency-domain result) 
clearly indicates that the influence of the momentum parameter $p$ 
on the waveform diminishes with increasing evolution time, in agreement with the theoretical expectation.

To summarize, we combine the results from both the frequency-domain and time-domain analyses, 
together with their mutual comparison, to draw the following conclusions:
\begin{itemize}
    \item \textbf{Frequency-domain perspective.} Perturbations as small as
    $\epsilon=0.001$ are sufficient to generate a complex QNM pair in the sampled
    double-barrier family. At this amplitude, the least-damped pole can remain purely
    imaginary while a more strongly damped complex branch is already present.
    Moreover, as the perturbation is located farther away from the brane,
    a larger number of additional modes emerge (see the left panel of Fig.~\ref{figure-7}).
    This qualitative feature is fundamentally different from the unperturbed case, 
    in which the QNM spectrum consists solely of purely imaginary modes. 
    \item \textbf{Time-domain perspective.} When the perturbation is either too weak or too far from the brane,
    a model-level oscillatory imprint remains in the numerical waveform. Its quantitative extraction becomes
    less robust as the amplitude is reduced or the perturbation is moved away from the brane
    (see Fig.~\ref{figure-8} and Table~\ref{Table-1}).
    By contrast, for an appropriate range of parameters 
    (namely, sufficiently large perturbation amplitudes and perturbations located closer to the brane),
    the effects manifest clearly during the intermediate stage of the waveform evolution.
    \item As discussed in Sec.~\ref{mechanism}, our estimate of the perturbation strength in the effective potential indicates 
    that perturbations located closer to the brane induce stronger deformations of 
    the effective potential (see Fig.~\ref{deltaVzvsdeltarhoz-am}). 
    Taken together with the results shown in the right panel of Fig.~\ref{figure-8}, 
    these results identify near-brane perturbations as the most favorable regime
    for producing a distinguishable oscillatory signature in the idealized waveform and hence define
    a concrete target for future source-to-detector modeling. They do not, by themselves, fix a
    detector-specific detection probability or parameter-recovery accuracy.
    This expectation relies on the fact that the unperturbed signal exhibits a purely decaying behavior, 
    whereas the presence of perturbations generates modes with nonzero real parts, 
    leading to oscillatory features that serve as model-level spectral signatures.
    On the other hand, it should be emphasized that the three-dimensional momentum $p$
    introduces a significant contamination in the early-time waveform, as illustrated in Fig.~\ref{figure-10}.
\end{itemize}

\section{Conclusion} \label{conclusion}

In this paper, we investigated the stability of QNMs in a dS braneworld background. 
Specifically, we modeled perturbations in the dS thick brane scenario by 
introducing a double-kink deformation into the background scalar field responsible for generating the brane.
By explicitly calculating the perturbed energy density distribution of the brane 
and the effective potential governing gravitational perturbations, 
we found that such matter perturbations induce a plateau-like uplift in the brane energy density (see Fig.~\ref{figure-2}), 
while generating a double-barrier-like deformation superimposed on the original effective potential (see Fig.~\ref{figure-2}). 
Furthermore, the ratio of the effective-potential variation to the energy-density variation indicates that
for perturbations of the same magnitude, deformations located closer to the brane ($z=0$) 
produce a significantly stronger impact on the effective potential.

To further assess how these deformations affect the QNM spectrum and the time-domain gravitational-wave signal,
we adopted an analytic superpotential proxy, parametrized by $(\epsilon,a)$,
to assess the spectral response.
Our frequency-domain analysis shows that, for the representative choice
$n=1/3$ and $\beta a=5$, the sampled value $\epsilon=0.001$---corresponding to
$\lVert\delta V^{(1)}\rVert_\infty/\lVert V_0\rVert_\infty\simeq1.5\times10^{-3}$---already
generates the complex pair $\omega/\beta\simeq\pm0.159502-0.513397i$.
The least-damped pole at this amplitude remains approximately
$\omega/\beta=-0.49988i$. This coexistence of a complex pair with the surviving
purely imaginary branch contrasts with the entirely imaginary unperturbed spectrum.
Moreover, the emergence and magnitude of the real parts depend sensitively on the perturbation location $a$,
as illustrated by the right panel of Fig.~\ref{figure-6} and by the
 location-dependent trajectories in Fig.~\ref{figure-7}.
We then examined how these newly generated modes with nonzero real parts manifest in the time-domain waveforms. 
We found that the early-time signal remains dominated by the unperturbed QNMs; 
however, the presence of a nonzero three-dimensional momentum $p$ can induce oscillatory behavior even during this early stage. 
At intermediate times, the effects of the perturbation become apparent, 
giving rise to a damped oscillatory phase governed by the perturbed QNM spectrum (see Fig.~\ref{figure-8}). 
Nevertheless, due to the dominance of the zero-mode sector over the simulated
late-time interval, this intermediate stage 
is relatively short, 
making it difficult to extract reliable information from time-domain signals when the perturbation amplitude is small.

This small effective deformation is consistent with the self-consistent
matter-sector reconstruction in Sec.~\ref{mechanism}. Equation~\eqref{Vrho-relation} and
Fig.~\ref{figure-4} show that, for the representative value $a_m=7/\beta$, the peak
tensor-potential response is only about $0.3\%$ of the peak energy-density change,
and that this response ratio decreases further as the matter structures move away from the
brane. Thus, a smooth matter-sector rearrangement can be transmitted to the tensor-scattering
problem as a substantially smaller potential deformation. Although $(\epsilon_m,a_m)$ and
$(\epsilon,a)$ are not related by a universal one-to-one map, this comparison provides a useful
reference scale: it shows how a visible matter-sector rearrangement can correspond to a much
weaker tensor-potential deformation. It therefore supports the physical plausibility of the
small effective barriers considered here without assigning $\epsilon=0.001$ to a unique
matter-density perturbation.

Consequently, the present dimensionless calculation establishes a controlled physical mechanism
and a robust model-level benchmark for future source-to-detector analyses. Within the first-order
effective model, the induced real parts generate a transient oscillatory stage whose amplitude
decreases as the deformation is weakened. Quantitative detector-level predictions---for example,
the detection probability, detectability with current instruments, constraints on the effective
four-dimensional cosmological constant, or parameter-estimation biases---require an
astrophysical source model and normalization, detector response, noise power spectral density,
and a parameter-estimation pipeline. These ingredients lie beyond the present calculation, but
their absence does not weaken the model-level conclusion: the dimensionless analysis establishes
the existence and parameter dependence of a physically motivated oscillatory QNM signature.
By identifying the amplitude and location regimes in which this normalized signature is enhanced,
it provides a reliable theoretical reference for future experimental and observational studies.

In future work, it would be of interest to extend the present analysis to other classes of perturbations 
and to investigate the stability of normal modes in AdS braneworld scenarios, 
which are directly related to corrections to the four-dimensional Newtonian potential and 
may have important implications for precision tests of gravity.
We also plan to perform a complete pseudospectral analysis of the deformed
operator with an explicitly specified operator norm, domain, compactification, and outgoing
boundary realization.

\section*{Data and Code Availability}

All Mathematica programs and numerical data supporting the Einstein--scalar reconstruction,
frequency-domain calculations, time-domain evolutions, and numerical error analysis are publicly
available in the GitHub repository cited in Ref.~\cite{Jia:2026data}. The repository preserves the
calculation hierarchy used in this work and includes the independent-method comparison and
convergence data reported in Table~\ref{table-qnm-method-comparison} and
Fig.~\ref{figure-qnm-convergence}.

\section*{Acknowledgments}

We would like to thank Wen-Yi Zhou for very useful discussions. 
This work was supported by 
the National Natural Science Foundation of China (Grants Nos. 12475056, 12205129, and 12247101),
the Fundamental Research Funds for the Central Universities (Grants Nos. lzujbky-2025-it05 and lzujbky-2025-jdzx07), 
the Natural Science Foundation of Gansu Province (Grants Nos. 22JR5RA389 and 25JRRA799), 
 Gansu Province's Top Leading Talent Support Plan, 
 and the ``111 Center'' under Grant No. B20063. 
 Wen-Di Guo was supported by ``Talent Scientific Fund of Lanzhou University''.
\par

\end{document}